\documentclass[preprint,12pt]{elsarticle}

\usepackage{amssymb}
\usepackage{amsmath}

\usepackage{algpseudocode}
\usepackage{algorithm}
\usepackage{mathtools}

\usepackage{graphicx}

\usepackage{hyperref}

\usepackage{xcolor}

\usepackage{caption,subcaption}

\journal{Internet of Things}

\begin{document}

\begin{frontmatter}



\title{Enhancing End-to-End Determinism and Reliability in 6TiSCH networks with disjoint leaf-based MPLS-like tunnels}


\author[inst1]{Lucas Aimaretto}

\affiliation[inst1]{
            organization={Facultad de Ciencias Exactas, Físicas y Naturales, Universidad Nacional de Córdoba},
            state={Córdoba},
            country={Argentina},
            }

\author[inst2]{Diego Dujovne}

\affiliation[inst2]{organization={Escuela de Informática y Telecomunicaciones, Universidad Diego Portales},
            state={Santiago},
            country={Chile}}

\begin{abstract}

Industrial multi-hop Internet of Things (IIoT) have strict reliability requirements and they are expected to have deterministic behavior. Reliability is associated with the network's ability to provide the best goodput possible to the destination from the source application, while deterministic behavior implies that the packets must also arrive at the destination before the maximum allowable deadline defined by the application expires. Although a relevant number of proposals have arisen in recent years, none of them achieve both restrictions simultaneously. In this work, we propose a cross-layer approach to solve this problem, by combining three strategies: (i) the use of the preferred parents (PP) and alternative parents (AP) together with the PRE (Packet Replication and Elimination) technique at the routing level; (ii) the use of MPLS tunnels from the leafNode, improving the Data Plane, to control the energy consumption and (iii) the use of the BDPC (Bounded Delay Packet Control) algorithm. The combination of the former strategies show that the behavior of the packet flows improves the end-to-end Packet Delivery Rate of the packets arriving before the deadline by 2.04 times with respect to standard Minimum Scheduling Function reference network while simultaneously increasing the minimum average network lifetime by 1.5 times, with respect to the hop by hop uncontrolled usage of PRE.

\end{abstract}


\begin{highlights}
\item Alternate Parent consideration in routing topology with distributed resource reservation.
\item MPLS-like source-based tunnels to reduce energy consumption.
\item BDPC scheduling function to control resource reservation as a function of the application's deadline.
\end{highlights}


\end{frontmatter}


\section{Introduction}
\label{section:intro}

Nowadays, industrial control systems are accelerating the pace of evolution towards the convergence of information technologies with operation technologies (IT/OT). This fact can be observed in the proliferation of an increasing number of measurement and control devices enabling real-time processing, storage and analysis systems to create new applications. Moreover, current IETF standards provide IPv6 support and interoperability to these devices.
 
Control and monitoring systems based on the Industrial Internet of Things (IIoT), unlike other type of applications, such as home or wearable applications, require a high PDR (Packet Delivery Rate) to the destination, considering only the packets arriving before a certain deadline. Furthermore, IIoT systems also requires a low and bounded energy consumption profile to enhance system autonomy and environmental sustainability. 

Deterministic behavior in a network becomes fundamental when the application data transported over the network is constrained by a maximum allowed time window, so as to keep the same behavior in time. For example, the power grid requires a deterministic telecommunications behavior in order to ensure that power lines can be activated within the required time threshold; in public transportation, deterministic behavior is used to guarantee that automated vehicles can be operated safely; in industrial control loops, deterministic behavior is used to reduce data loss which mey trigger an unexpected stop in an industrial process due to a lack of proper monitoring; or even in the entertainment industry determinism is used to enable Audio/Video Bridging (AVB) or to decrease the cost of maintenance in public attractions. \cite{thubert2017converging, koutsiamanis2018best, draft-ietf-raw-use-cases}.

An ideal deterministic flow is predicable across a network, without any kind of interference or influence from other flows. A deterministic network can transport different flows which must meet their reliability and determinism goals without interference between them. Current internet QoS techniques cannot provide or ensure this kind of behavior, since QoS implementations enhance the performance of one flow to the detriment of the other concurrent flows \cite{thubert2017converging, jenschke2020toward}.

Therefore, a data flow is deterministic when each packet has maximum arrival time restriction: packets arriving later than the time limit become useless and are considered lost thus reducing the application-level Packet Delivery Rate (PDR). A standard (non-deterministic) network must be configured and adjusted to enable deterministic flows to comply with PDR and deadline application constraints \cite{AIMARETTO2023100778}.


Among the currently available multi-hop IIoT network solutions,   the most efficient and flexible standardized technology is Time Slotted Channel Hopping (TSCH), where the PHY and MAC layers are defined in the IEEE 802.15.4 standard \cite{ieee2015ieee}. TSCH defines a time-channel matrix where each cell located by the coordinates \texttt{[slotOffset,channelOffset]} establishes a communication opportunity between nodes to exchange the packets. 

\begin{figure}
	\centering
	\includegraphics[scale=0.6]{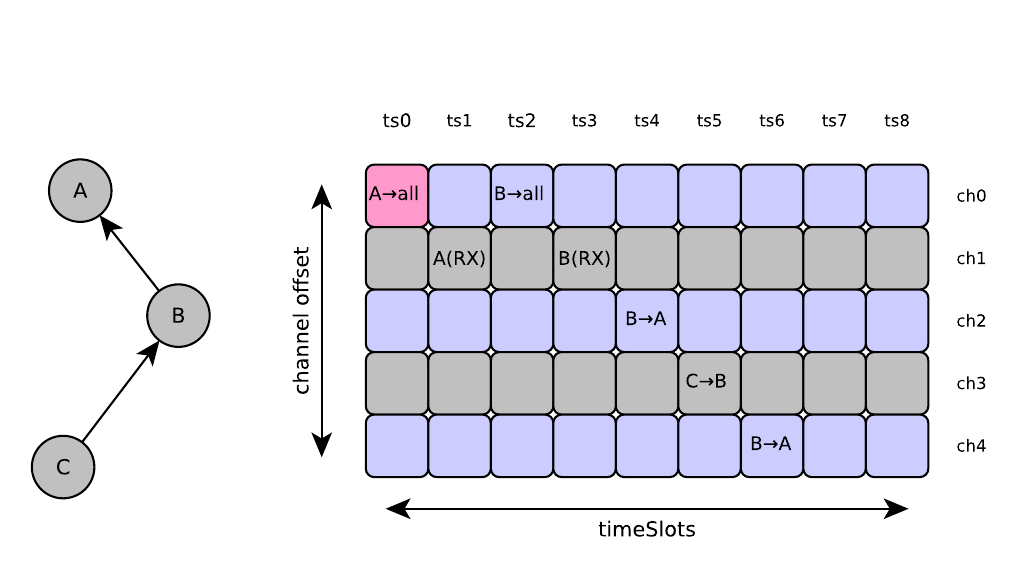}
	\caption{TSCH slotFrame. The schedule programmed on the slotFrame instructs the node the time to wake up or go to sleep. The cell on coordinate 0x0 is called the MinimalCell and is used for broadcast traffic. Source: \cite{AIMARETTO2023100778}.}
	\label{fig_tsch_slotFrame}
\end{figure}

In a TSCH network, all nodes participating in the network must build a schedule where they agree on which cells they will use to exchange packets to their neighbors, turning on the receiver during the cell period and - if there is a packet waiting - transmitting in that same period. The rest of the time, the transmit and receive stages of the nodes not participating in that communication opportunity remain off. Therefore, TSCH requires that all participating nodes are synchronized. An example of a network with three nodes and a schedule defined in the TSCH matrix is shown in Fig. \ref{fig_tsch_slotFrame}. In the example, node B will transmit to node A when timeSlot 4 occurs, on channelOffset 2.

\begin{figure}
	\centering
	\includegraphics[scale=0.6]{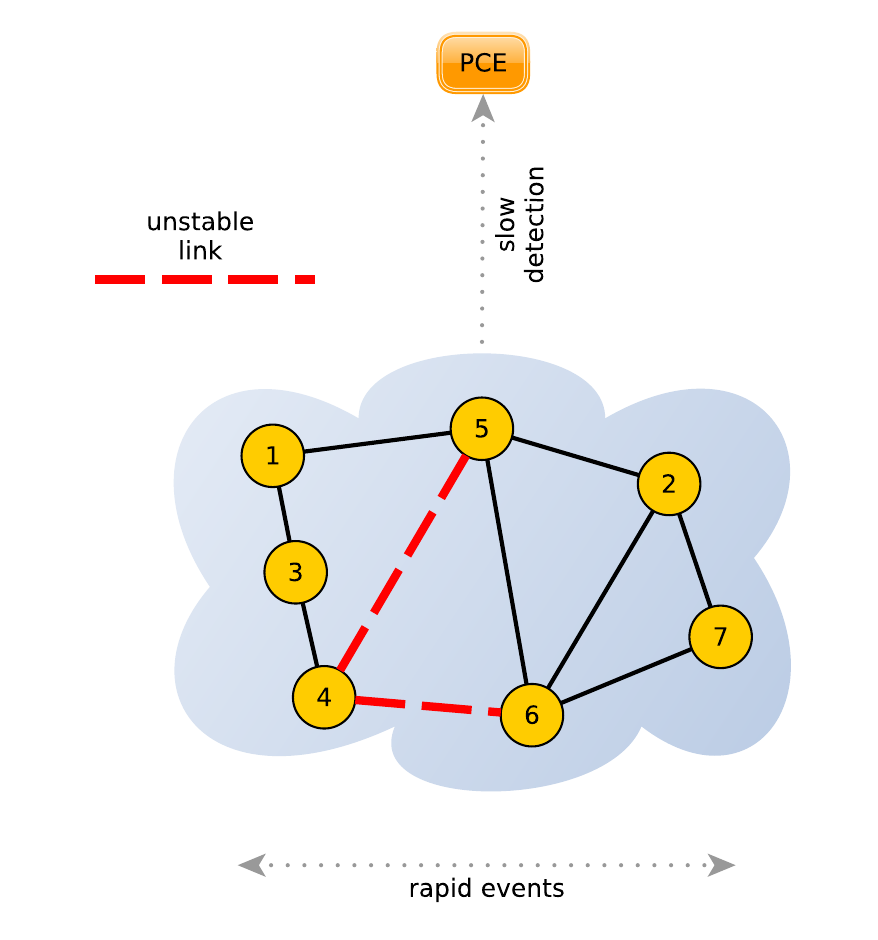}
	\caption{Location of the Path Computation element (PCE) in a mesh network. The PCE is not really suitable for wireless meshed networks as it cannot react at the same pace as events do develop in the network.}
	\label{fig_raw_pce}
\end{figure}

The construction of the 6TiSCH schedule can be carried out using different methodologies, which can be classified between centralized and distributed. Centralized scheduling algorithms can be fixed or dynamic. In a centralized scheduling algorithm, resource allocation is performed by the Path Computation Element (PCE), which is an external module which controls the network schedule distribution. The PCE receives statistics from the nodes and returns instructions on how to build the schedule by applying an internal algorithm. The PCE calculates and installs the paths between nodes and provides information to guide the routing decision. However, in a wireless network, the management packets path to and from the PCE tends to have high energy consumption and low PDR, thus reducing the ability to react to link changes occurring in the network with in a timely manner \cite{urke2021survey,draft-ietf-raw-architecture}. The operation of the PCE in the context of a wireless network can be observed in Fig. \ref{fig_raw_pce}.

The alternative to centralized scheduling is distributed scheduling, where each node takes scheduling decisions and negotiates with its neighbors how to build a shared schedule. This is achieved through Scheduling Functions (SFs), which are part of the IPv6 over TSCH mode of IEEE 802.15.4e (6TiSCH) stack \cite{rfc9030}. SFs manage resource allocation with the aim of improving the performance of specific network characteristics such as delay, energy, network lifetime and packet delivery, among others. In an industrial context, the most relevant network performance characteristic is deadline-aware PDR, which accounts only for the packets arriving before the application deadline. In addition, given the context of sustainability and maintenance requirements, the solution must be energy efficient. However, the current standardized SF, called Minimal Scheduling Function (MSF) \cite{rfc9033}, is a distributed function that allocates resources in the schedule based on traffic demand and was designed specifically for Best-Effort traffic with occasional traffic peaks. Using MSF, two neighboring nodes negotiate capacity in the child$\rightarrow$parent direction using the 6P \cite{rfc8480} protocol.

Network robustness and reliability is also responsibility of the end-to-end path construction generated by the routing protocol. The default routing protocol in this type of network, IPv6 Routing Protocol for Low-Power and Lossy Networks (RPL) \cite{rfc6550}, constructs a Destination Oriented Directed Acyclic Graph (DODAG): A DODAG is a destination-oriented tree topology. The destination, or root, is simply the central node to which all other nodes in the network send their data. The root of the DODAG is an edge router that may have a connection to other networks or even the Global Internet.

\begin{figure}
	\centering
	\includegraphics[scale=0.6]{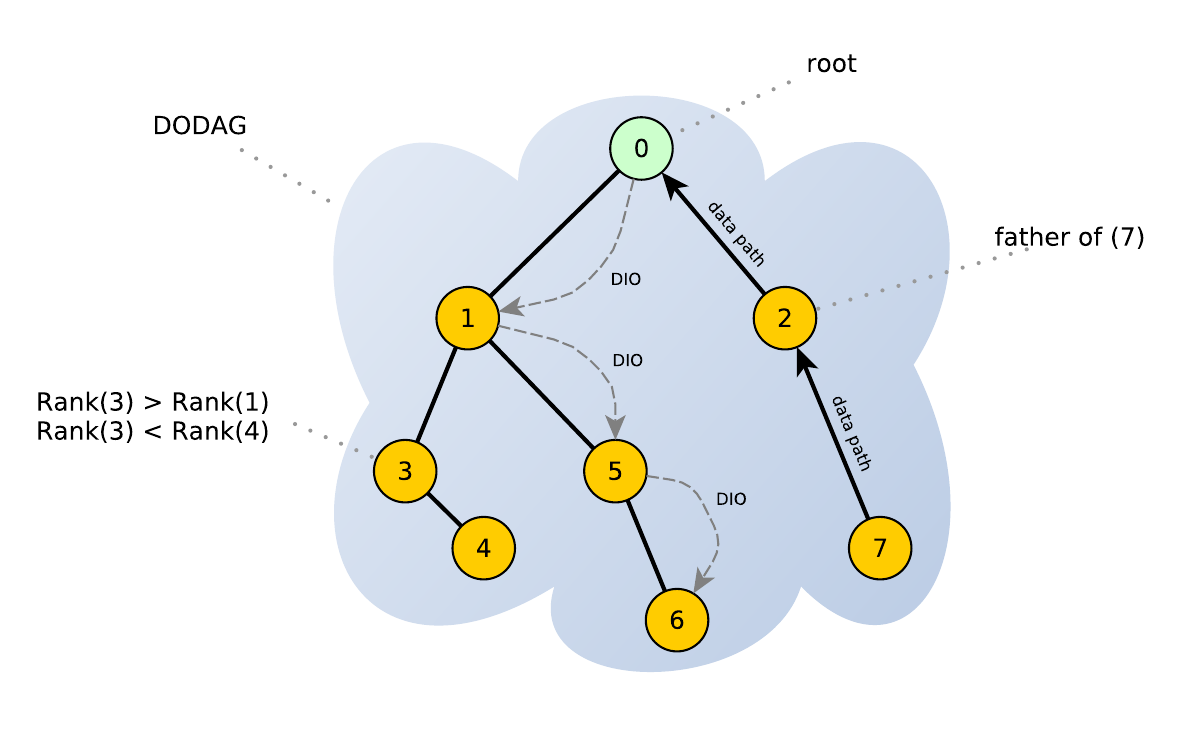}
	\caption{RPL forms a so called Destination Oriented Directed Acyclyc Graph (DODAG). The Rank helps a node identify its topological position within the network. The smaller the Rank, the closer to the root a node is. Source: \cite{AIMARETTO2023100778}.}
	\label{fig_rpl_dodag}
\end{figure}

RPL builds the DODAG following an attribute called Rank. Each participating node in the network has a Rank value, which is assigned when the node joins the network. The Rank value is announced among neighbors using protocol messages called DODAG Information Object (DIO). When a node receives a DIO packet from a neighbor, it registers the Rank value in an internal list, to select a parent node among them. Once the network has completed the convergence process, each node then registers with the preferred parent (PP), which will be the default next-hop when sending traffic to the root node. A schematic of a DODAG with the root node and the routes between nodes can be observed in Fig. \ref{fig_rpl_dodag}.

In a 6TiSCH network \cite{rfc9030}, the TSCH, MSF and RPL protocols must work harmonically. The RPL protocol periodically sends DIOs messages so that the nodes can choose the parent with the lowest Rank value among the available neighbors to send packets to the DODAG root. However, during the lifetime of the network, a parent change event may occur, as a consequence of a change in the Rank value of one or more neighbors. In this case, the node must restart the parent selection process and the node negotiates cells with the new parent using the 6P protocol to reassign the existing cells in the child$\rightarrow$parent$_{old}$ link to the child$\rightarrow$parent$_{new}$ link. From this point on, MSF starts monitoring the child$\rightarrow$parent$_{new}$ link to adjust the number of assigned cells assigned, depending on the traffic demand.

Although current 6TiSCH networks have a better deterministic behavior than competing standardized proposals based on CSMA or modified ALOHA, this feature is still not enough to comply with for IIoT network requirements. IIoT networks must meet real-time requirements also. For a robust and predictable end-to-end behavior in an IIoT network, transmitted packets shall not be discarded or lost along the path to the DODAG root, and they must arrive at the destination before the application-defined deadline.  A data packet that arrives at its destination later than the deadline is considered lost, and must be discarded by the application \cite{AIMARETTO2023100778,ryu2005urgency}, thus affecting the  application performance.

In this paper, we present a solution where different path-to-root redundancy strategies are combined to reduce packet loss, along with the application of BDPC \cite{AIMARETTO2023100778} to increase the number of packets arriving before the application $deadline$ at the destination. Consequently, the contributions of this work are:



\begin{itemize}
    \item To enable the use of an alternate parent (AP) in a network with distributed resource allocation;
    \item To reduce power consumption by enabling disjoint paths for data flows based on MPLS mechanisms at the leaf-node.
    \item To leverage BDPC resource allocation to overcome link and node variations in packet flow capacity and deliver critical packets to the destination before a maximum application $deadline$.
\end{itemize}


The rest of the paper is organized as follows: Section \ref{stateArt} presents the current state of the art; Section \ref{section:situacion} presents the development of our solution with a deeper analysis of the contributions; Section \ref{section:simul_setup} presents the simulator with the implementation; Section \ref{section:simul_results} presents the results; and finally, Section \ref{section:conclusion} concludes this work.
\section{State of the art}\label{stateArt}

Path diversity increases the probability of packet arrival at the destination. In 6TiSCH-based IIoT networks, this fact implies that the RPL protocol must be able to use more than one path to the destination. Jenschke et al.\cite{jenschke2019alternative}, analyze the possibility of using an alternate parent (AP) in addition to the preferred parent (PP) to increase the redundancy of the Control Plane, so as to generate multiple paths for the Data Plane.

In the standard version of RPL, a node within a DODAG manages a group of neighbors, which are possible candidates to become a preferred parent (PP). This group of neighbors is referred as the Parent Set (PS). Jenschke et al.\cite{jenschke2019alternative} propose three different configuration approaches between the PP and the AP, with respect to what is called Common Ancestor (CA): Any node can be considered an alternative parent only if it has some Common Ancestor (CA) with the current Preferred Parent (PP). Consequently, it turns out to be necessary that DIO messages also share information about common ancestors, as specified in the document \cite{draft-ietf-roll-nsa-extension-10}. As a matter of fact, any node in the IIoT network can have an alternate parent (AP) based on the following configurations:

\begin{itemize}
	\item \emph{strict} CA: to elect a node as AP, its PP must be equal to the PP of the current preferred parent.
	\begin{itemize}
		\item $PP(PP) = PP(AP)$
	\end{itemize}
	\item \emph{medium} CA: to choose a node as an AP, its PP must belong to the Parent Set (PS) of the current preferred parent.
	\begin{itemize}
		\item $PP(AP) \in PS(PP)$
	\end{itemize}
	\item \emph{soft} CA: to choose a node as an AP, its parent set must have a node in common with the parent set of the PP.
	\begin{itemize}
		\item $\exists \: [PS(AP) \cap PS(PP)]$
	\end{itemize}
\end{itemize}

\begin{figure}
	\centering
	\includegraphics[scale=0.6]{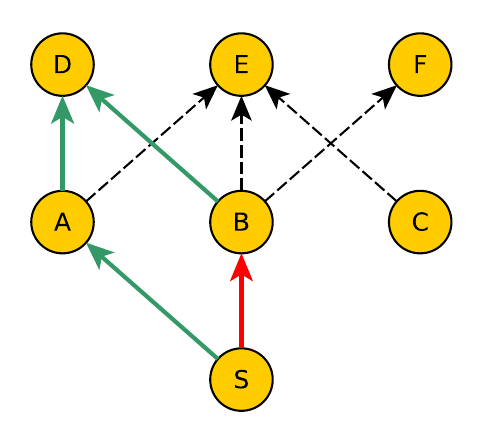}
	\caption{The \emph{strict} method to select an alternate parent (AP) implies that the $PP(PP) = PP(AP)$. In this example, $A$ is the PP of $S$. Because $PP(A) = PP(B)$, then $B$ is the AP for $S$.}
	\label{fig_rpl_ap_strict}
\end{figure}

An example of AP selection using the strict approach is shown in Fig. \ref{fig_rpl_ap_strict}.


If the AP becomes part of the DODAG, the resulting improvement has effect only at the Control Plane level: The RPL routing process enables an additional next-hop for packets traveling to the DODAG root. In this proposal we enable also the Data Plane to take advantage of the use of APs.

\begin{figure}
	\centering
	\includegraphics[scale=0.6]{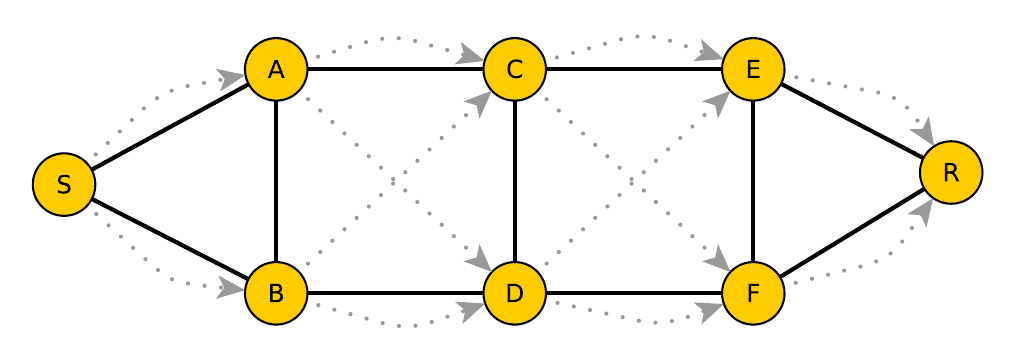}
	\caption{Packet Replication and Elimination (PRE). The data packets    are copied along the network hop-by-hop, in their path towards their destination. Once the first packet of a given flow arrives to $R$, the rest will be dropped on their arrival.}
	\label{fig_raw_pre}
\end{figure}

One strategy for the Data Plane to take advantage of the additional next-hop offered by the Control Plane is to implement the Packet Replication and Elimination (PRE) \cite{draft-ietf-raw-architecture} technique. With this method, a node sends a copy of the received packet to more than one parent in separate transmissions. For example, in Fig. \ref{fig_raw_pre}, the source node $S$ wants to send packets to the remote destination $R$. $S$ generates a copy of the data packet and sends it to different parents. This behavior is repeated hop by hop. Once the first packet arrives at the destination, the root will discard the copies arriving later. However, this process cannot be unbounded: The constant generation of packet copies without any hop-by-hop control generates high energy consumption \cite{giorgiosPacketDuplicationEnergy}.

According to the results analyzed in \cite{jenschke2019alternative}, when only one preferred parent (PP) is available, the $PDR_{e2e}$ is 82.7\%. To the contrary, the use of an AP and the PRE packet replication technique raises the $PDR_{e2e}$ to 97.32\% using a \emph{strict} approach; to 99.66\% for the \emph{medium} case and 99.98\% for the \emph{soft} case. However, the solution proposed in \cite{jenschke2019alternative} is based on a centralized and static scheduling. In wireless IoT networks, the usage of centralized resource management is often discouraged because of the inability of the Path Computation Element (PCE) to react to the high variability of the wireless network \cite{urke2021survey,draft-ietf-raw-architecture}.


Koutsiamanis et al. \cite{giorgiosLFC} propose the Leap Frog Collaboration (LFC) strategy. LFC also uses the PRE technique, which deletes packets received by intermediate nodes, only when the node has previously received a first copy of the packet. However, the generated schedule is also static, centrally managed and manually constructed: ``The schedule is statically defined and pre-calculated offline for the given topology''. The results of \cite{giorgiosLFC} show that the $PDR_{e2e}$ is better than 99.83\% for all simulations performed with $PDR_{link}\geq80\%$.

Another proposed strategy to increase the efficiency of the Data Plane is the use of Multi Protocol Label Switching (MPLS) \cite{rfc3031} labels to construct paths in the network. In \cite{Morell2013Label}, by using MPLS tunnels, signaled by RSVP \cite{rfc2205}, the reliability of the data plane is increased by also improving the end-to-end delay distribution and the total network throughput, reaching up to 70\% of the theoretical maximum. However, in \cite{Morell2013Label}, the ability to use an alternative parent to forward data packets is not considered. Therefore, if a failure of the preferred parent occurs, the routing protocol will have to start a new convergence process to find a new candidate, introducing delays in the packet flow that are not acceptable for an IIoT application.

Another important issue in the industrial context is the delivery of critical packets to the destination within the maximum time allowed \cite{AIMARETTO2023100778}. The $PDR_{e2e}$ details the rate of packets arriving successfully at the destination, $PDR_{e2e}=n_{rx}/n_{tx}$. An acceptable industrial network should have a $PDR_{e2e}\geq99.9\%$. However, a critical packet that arrives at the destination later than the stipulated deadline must be discarded by the application and therefore reduces the $n_{rx}$ number affecting the final $PDR_{e2e}$ value.

A new scheduling function, called Bounded Delay Packet Control (BDPC), is presented in \cite{AIMARETTO2023100778}. BDPC proposes a different strategy for resource reservation. Based on the time budget that a data packet has consumed, cells are allocated to the slotFrame in the parent$\rightarrow$child direction. Aimaretto et al. \cite{AIMARETTO2023100778} show that using this technique, the number of delivered packets before the application deadline improves up to 2.6 times compared to the Minimal Scheduling Function (MSF), for the same network, same application and identical conditions.

\begin{figure}
	\centering
	\includegraphics[scale=0.6]{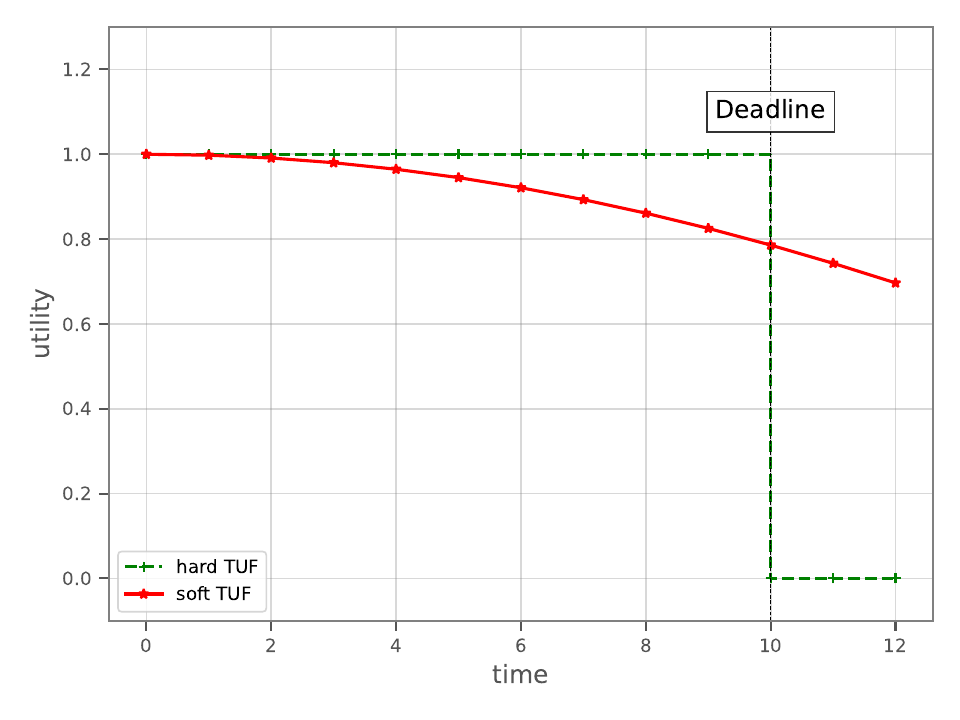}
	\caption{Hard and Soft Time Utility Functions. For the hard case, if the task is completed after the maximum deadline, the outcome is discarded, even if this outcome is correct. Source: \cite{AIMARETTO2023100778}.}
	\label{fig:tuf}
\end{figure}

BDPC builds its operating logic on the basis that critical packets must reach their destination before a maximum allowed deadline. There are two types of \emph{deadline}: hard and soft. The hard case defines the maximum limit in which the task must be completed, whereas on the soft deadline, the task permits variability at the end time of a task. The hard case is associated to real-time applications, which include industrial control applications. A real-time task means that the task must be completed within a maximum time limit, otherwise it would no longer be useful.  Examples of Time Utility Functions (TUF) can be seen in Fig. \ref{fig:tuf}.

Aimaretto et al. \cite{AIMARETTO2023100778} proposes a new parameter called $latePaqs_{e2e}=n_{delayed}/n_{rx}$. $latePaqs_{e2e}$ indicates the rate of late packets arriving at the root node, considering the end-to-end path. Therefore, the $1-latePaqs_{e2e}$ value measures the rate of packets that arrived within the maximum $deadline$ supported by the application. $latePaqs_{e2e}$ is related to the deadline as follows:

\begin{equation}
    \label{latePaqs}
    1-latePaqs_{e2e}=f(deadline)
\end{equation}

\bigskip

In an industrial context, late packets arriving at the destination become discarded. Therefore, in order to improve $1-latePaqs_{e2e}$, BDPC creates a local variable called $latePaqs_{link}$. This parameter measures the rate of late packets arriving at an intermediate node and it is used by Alg. \ref{alg:slots}, which manages the cells in the parent$\rightarrow$child direction. Based on the comparison of $latePaqs_{link}$ against the \texttt{sfMax} and \texttt{sfMin} variables, BDPC generates the allocation of resources in the parent$\rightarrow$child direction. There is a $latePaqs_{link}$ value for each child node, where each child is identified by its MAC address. Consequently, BDPC is agnostic to the network topology.

The \texttt{sfMax} value sets the upper limit above which Alg. \ref{alg:slots} adds cells to its child. As a result of this action, the work in \cite{AIMARETTO2023100778} results in:

\begin{equation}\label{bdpc_prob}
    p(delay\leq deadline) = 1-\texttt{sfMax}
\end{equation}

\begin{algorithm}
\caption{BDPC: cell assignment to the PreHop}\label{alg:slots}
\begin{algorithmic}
\medskip
\State $latePaqs_{link} \xleftarrow[]{get} DB_{data}$

\medskip
\If{$latePaqs_{link} \geq \texttt{sfMax}$}
    \State $parent \xrightarrow[]{6P}addCell(child)$
\ElsIf{$0 \leq latePaqs_{link}$ and $latePaqs_{link} \leq \texttt{sfMin}$}
    \State $parent \xrightarrow[]{6P}delCell(child)$
\EndIf
\end{algorithmic}
\end{algorithm}

\section{Proposed Solution}

\label{section:situacion}

In this paper, we propose a solution to the data transmission requirements in IIoT networks with real-time constraints by combining simultaneous improvements between the control plane and the data plane with a cross-layer approach. We designed and implemented a set of new algorithms for RPL, MSF and 6TiSCH protocols, in order to control the behavior of the control plane and the data plane respectively.

In RPL and MSF, we propose to increase the next-hop diversity along the path to the root node to improve the Control Plane behavior, while we propose to use BDPC as SF to fulfill the application $deadline$ for 6TiSCH. Finally, we incorporate the use of MPLS tunnels to improve the reliability of the Data Plane.

\subsection{Minimal Scheduling Function (MSF)}

The Minimal Scheduling Function (MSF) \cite{rfc9033} is the standard SF included within the 6TiSCH protocol stack. It was originally designed to comply with best-effort traffic requirements. MSF is a distributed SF that manages cells in the slotFrame in the child$\rightarrow$parent direction, depending on the traffic load. Alg. \ref{alg:msf} describes how MSF works: if the traffic load, determined by the Negotiated Cells Used ($NCU$) variable is higher than the $Lim\_High$ threshold, a 6P request is triggered towards the preferred parent (PP) and an extra cell is added to the schedule. If instead, $NCU < Lim\_Low$, a 6P request is triggered towards the preferred parent (PP) and an extra cell is removed from the schedule. If the traffic load remains between the two thresholds, no action is taken. This mechanism is repeated when a number of cells defined by the Negotiated Cells Elapsed ($NCE$) variable elapses.

\begin{algorithm}
\caption{MSF: cell assignment to the NextHop}\label{alg:msf}
\begin{algorithmic}
\medskip
\State $NCE =$ Negotiated Cells Elapsed
\State $NCU =$ Negotiated Cells Used
\State $Max\_NumCells= 100$
\State $Lim\_High= 75$
\State $Lim\_Low= 25$

\medskip
\If{$NCE > Max\_NumCells$}
    \If{$NCU > Lim\_High$}
        \State $child \xrightarrow[]{6P}addCell(PP)$
    \ElsIf{$NCU < Lim\_Low$}
        \State $child \xrightarrow[]{6P}delCell(PP)$
    \EndIf
\EndIf
\end{algorithmic}
\end{algorithm}

MSF manages the number of cells assigned to its PP according to the traffic load. The default values $Max\_NumCells=100$, $Lim\_High=75$ and $Lim\_Low=25$, correspond to what is defined in the standard \cite{rfc9033}. 

MSF considers the possibility of a parent change. When this happens, the following three steps are fulfilled:

\begin{itemize}
    \item The child node counts the number of negotiated cells assigned to the current PP$_{old}$.
    \item The child node triggers one or more 6P \texttt{ADD} commands to the new parent, PP$_{new}$, proposing the same number of existing negotiated cells which enabled the data path to the PP$_{old}$.
    \item When signaling is completed, a 6P \texttt{CLEAR} message is sent to PP$_{old}$, to release the negotiated cells.
\end{itemize}

However, when path diversity is incorporated, when an alternate parent (AP) is available, the reference in the child node must change. According to \cite{rfc9033}, $NCE$ and $NCU$ counters are counters maintained by a node with respect to its current PP$_{current}$. When the AP is added as an alternative, the counters are individualized in a per-neighbor basis and not necessarily a single PP$_{current}$. This is particularly important in a situation of parent change: a child node now may have two or more alternative parent nodes and in case of using MSF, the node needs to implement independent $NCE$ and $NCU$ counters per parent.

\subsection{Improvement in the Data Plane}\label{flooding}

After the convergence process in the Control Plane, where the preferred and alternate parents have been selected by the RPL routing protocol with AP support, the Data Plane can start transmitting packets, using the available next-hops towards the root.

The Packet Replication and Elimination (PRE) \cite{draft-ietf-raw-architecture} technique is one of the possible alternatives for packet management in networks with a diversity of paths to the root. Since the PRE technique involves generating copies of the same packet, the root node will choose the first one to arrive by discarding subsequent copies. In order for a router to detect whether it has already received a copy of a packet, we define a data flow.

\paragraph{\textbf{Flow}} A data flow is defined as the succession of packets where each packet contains the source IP fields of the \emph{leafNode} that originally generated each packet plus a sequence number that increases with the generation of new packets. Formally, a flow is then composed of a series of tuples, namely: $[(\text{IP}_{j},seq_{0}),(\text{IP}_{j},seq_{1}),(\text{IP}_{j},seq_{2}),...,(\text{IP}_{j},seq_{n})]$, where $n$ is the last packet created by the \emph{leafNode} so far. The sequence number of the packet is assigned at the time of its creation at the \emph{leafNode}.

\bigskip

For a tuple $(\text{IP}_{j},seq_{m})$ with $0 \leq m \leq n$ that is received by an intermediate node, if it is detected as repeated, the node can take action, either (i) generate a copy of the packet and send it to a different parent; (ii) forward the packet without further action; (iii) discard the packet.

\begin{figure}
	\centering
	\includegraphics[scale=0.5]{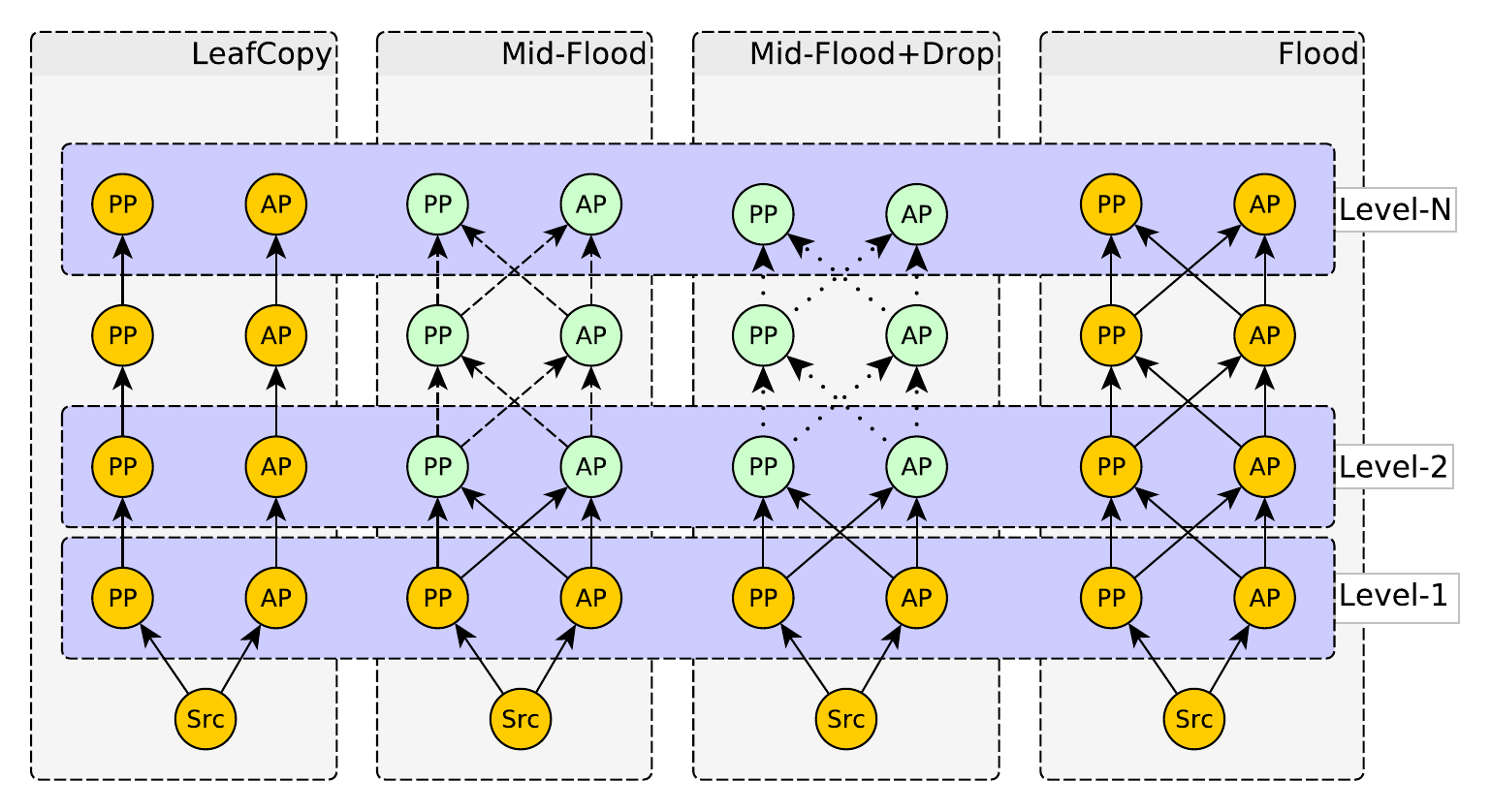}
	\caption{Different kind of \emph{flooding} in a network with an alternate parent. The figure shows the data plane, and the solid arrows show the data path of the replicated data packets. In the case of Mid-Flood and Mid-Flood+Drop, the dashed lines represent that received packets will be copied and forwarded only if that specific flow has never been received before.}
	\label{fig_ap_flood}
\end{figure}

Fig. \ref{fig_ap_flood} shows different packet copying strategies in the Data Plane. The management of the Data Plane is based on the use of labels which are inserted in the data packet for the first time when it was created at the \emph{leafNode}. These labels indicate which \emph{next-hop} will be used by the packet as the forwarding relay, either the PP or AP. Consequently, the valid labels are either \texttt{"PP"} or \texttt{"AP"}. A tunnel entity starting at the \emph{leafNode} refers to the source-routing concept, i.e., the definition of the paths via the corresponding parents is done at source. The next step is the forwarding stage in the router. The source node will send each copy to the appropriate next-hop, according to what is indicated in the label of each copy of the packet. 

This labeling technique is used, for example, in Multiple Protocol Label Switching (MPLS) \cite{rfc3031}, where the label inserted in the data packet indicates which next-hop to use, whereas the final destination is indicated by the packet's destination IP. Labels are actually another way of visualizing a Label Switched Path (LSP).

\begin{figure}
	\centering
	\includegraphics[scale=0.4]{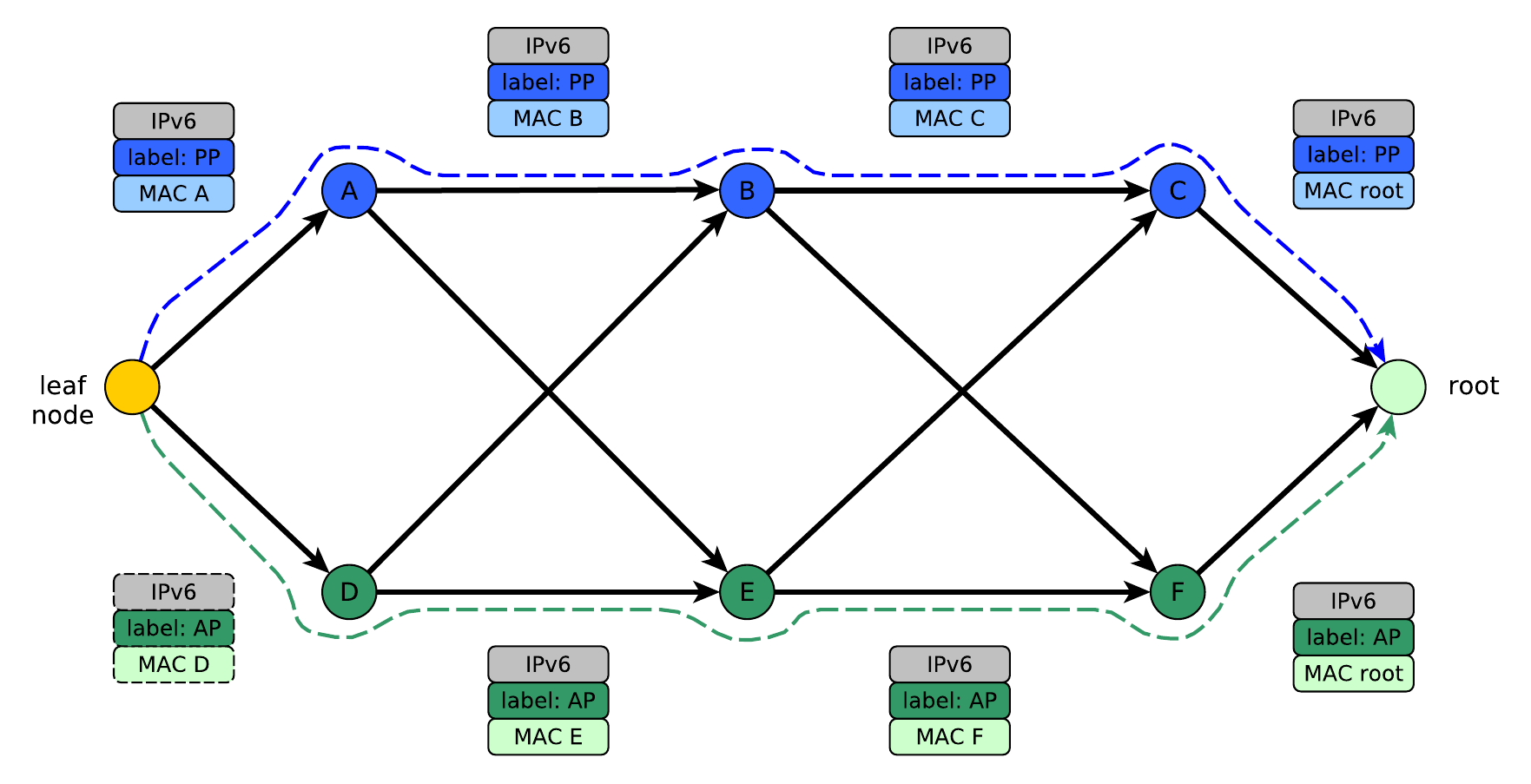}
	\caption{Data plane with LSPs signaled at the source with a label, which defines the path to the destination via the PPs and APs. This is the \emph{leafCopy} technique: two independent topologies are defined based on the labels assigned at the source. The blue nodes are the PPs; the green nodes are the APs. The label remains unchanged until the destination. In this case, the different LSPs created at source can be seen.}
	\label{fig_sr_leafCopy}
\end{figure}

\begin{figure}
	\centering
	\includegraphics[scale=0.4]{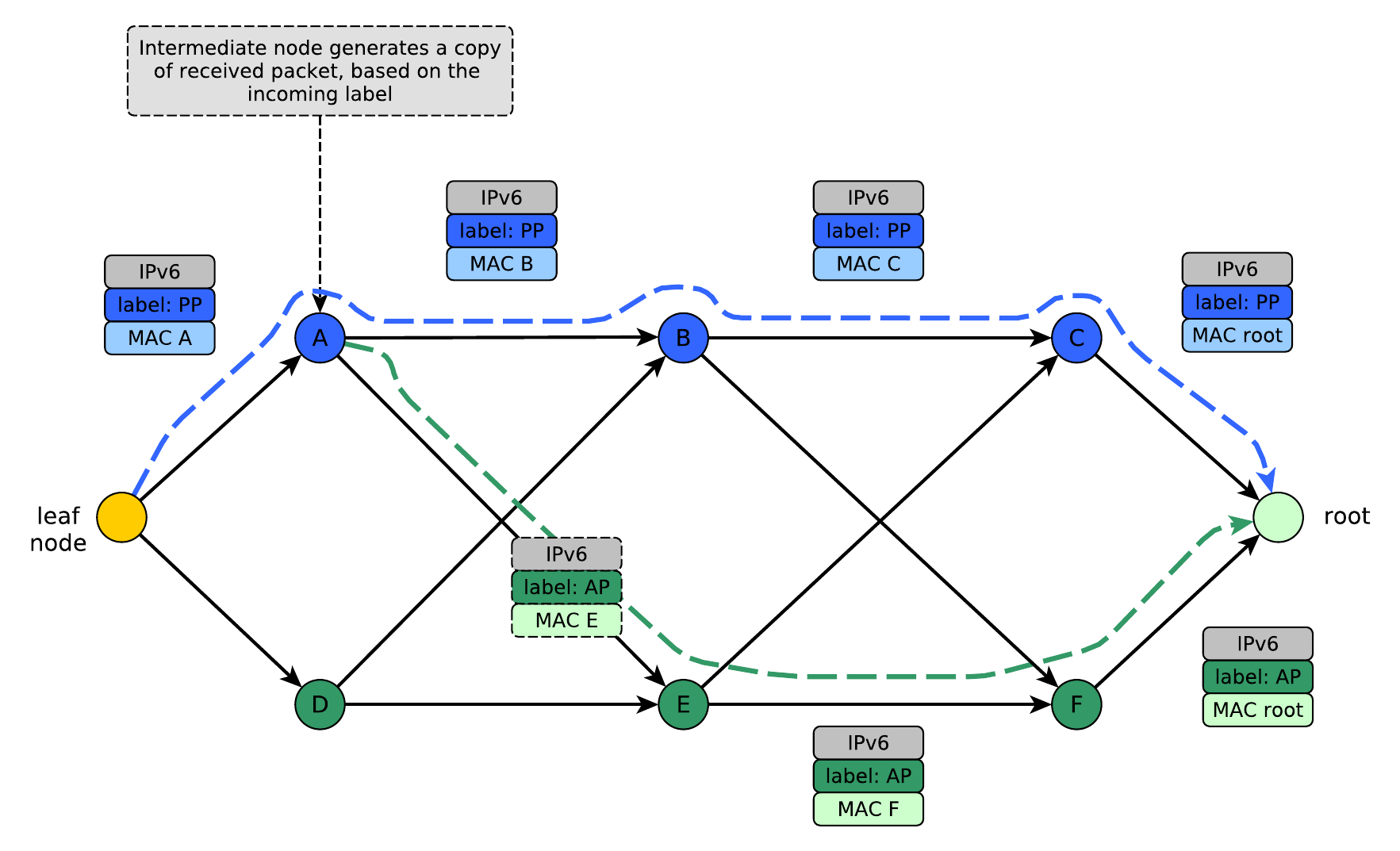}
	\caption{Packet replication at the intermediate node. An intermediate node will generate copy of the received packet based on the incoming label. Only applicable to the approaches \emph{mid-flood}, \emph{mid-flood-drop} and \emph{flood}.}
	\label{fig_sr_copy}
\end{figure}

The nodes will react to the packets received from their children according to the $flooding$ strategy, as shown below:

\begin{itemize}

	\item \textbf{LeafCopy}: Copies are only generated at the source node and sent to each parent. The parents, once they receive the traffic from their child, analyze the packet label and switch the packet to the corresponding parent. This is shown in Fig. \ref{fig_sr_leafCopy}.
	
	\item \textbf{Mid-Flood}: The source node generates copies of each data packet and forwards each copy to the corresponding parents. Each parent, once the packet is received, accounts for the flow and only generates copies of the received packet if the corresponding tuple $(\text{IP}_{j},seq_{m})$ has not passed through such a node before. If a copy of the received packet (or tuple) has already passed through, the received packet is switched to the corresponding parent by honoring the original label of the packet, without making new copies. Fig. \ref{fig_sr_copy} shows an example of copying at an intermediate node.
	
	\item \textbf{Mid-Flood+Drop}: The source node generates copies of each data packet and forwards each copy to the corresponding parents. Each parent, once the packet is received, takes account of the flow and only generates copies if the received packet has not passed through that node before. If the tuple $(\text{IP}_{j},seq_{m})$ has already passed through such a node, no copies or forwarding is done and the packet from the child is discarded.
	
	\item \textbf{Flood}: The source node generates copies of each data packet and forwards each copy to the corresponding parents. Each parent, once the packet is received, re-generates copies of the packet regardless of whether the flow has already passed through.
	
\end{itemize}

In either case, the root node keeps the first packet received, and discards the rest of the copies that are received later.

\subsection{Switching with absent parent}

In Section \ref{flooding}, we discussed how to generate an LSP at the source to send traffic from the leafNode to the root. Depending on the type of \emph{flooding}, packet copies will be generated to maximize the $PDR_{e2e}$. The root node will choose the copy that reaches the destination first, discarding subsequent copies. The copies are identified according to the data flow using the $(\text{IP}_{j},seq_{m})$ tuples.

In this paradigm, all packets carry a label that indicates via which next-hop an intermediate node should switch the packet on the path to the root. Fig. \ref{fig_sr_leafCopy} shows the case \emph{leafCopy} where the source node generates the copies, applies the labels, and based on the labels, decides to send each copy to each of the available parents (either AP or PP).

In the case the link or node becomes unstable, the router decides how to forward the received packet based on the packet label and the  availability of the parent nodes.

\begin{figure}
	\centering
	\includegraphics[scale=0.4]{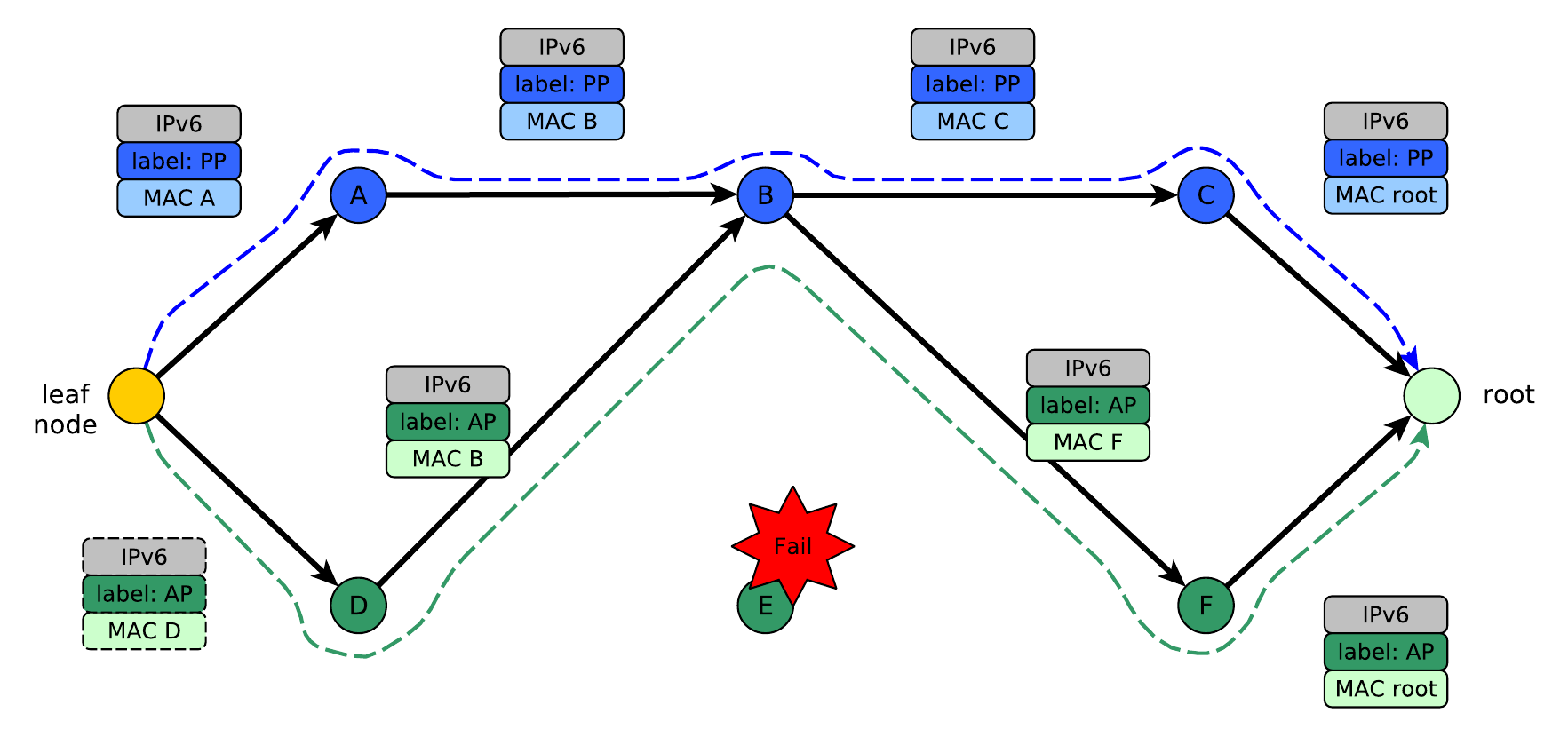}
	\caption{A data plane with a path signaled at source with a label. An intermediate node, upon receiving a packet, analyzes the label and then switches based on the label. If this is not possible, the label information is preserved in the forwarded packet and the next node in the path with enough parent alternatives makes the required switchover.}
	\label{fig_sr_leafCopy_problems}
\end{figure}

This situation can be seen in Fig. \ref{fig_sr_leafCopy_problems}. Router D receives a packet from its child, a \emph{leafNode}, whose label is \texttt{"AP"}; therefore, D should send the packet via $AP(D)=E$. Since D's AP is not available, D proceeds to send the packet to the available parent, which in this case is $PP(D)=B$. Then, router B receives a packet from D, whose label is \texttt{"AP"}. Since B does have an AP available, then it forwards the packet to the AP. Consequently, the packet follows the LSP signaled at the source along the path to the destination.

Alg. \ref{alg_chap5_mac_selection} describes the procedure for choosing the MAC address in order to switch the packet. If, for any reason there are no parents available, the MAC address will be null and the packet will be discarded at the intermediate node switching stage.

\begin{algorithm}
        \floatname{algorithm}{Algorithm}
	\caption{Hop-by-Hop MAC selection during forwarding }\label{alg_chap5_mac_selection}
	\begin{algorithmic}
	\medskip
        \State $fwdPacket \xleftarrow[]{copy} rxPacket$
        \State $label \xleftarrow[]{get} rxPacket_{label}$
        \State $PP \xleftarrow[]{get} router_{PP}$
        \State $AP \xleftarrow[]{get} router_{AP}$

        \medskip
 
	\If{$label = \texttt{"PP"}$}
            \medskip
            
            \If {$PP \text{ is } True$}
                \State $fwdPacket_{MAC}=PP_{MAC}$
            \ElsIf {$AP \text{ is } True$}
                \State $fwdPacket_{MAC}=AP_{MAC}$
            \Else
                \State $fwdPacket_{MAC}=None$
            \EndIf
            
            \medskip
	\ElsIf {$label = \texttt{"AP"}$}
            \medskip
            
            \If {$AP \text{ is } True$}
                \State $fwdPacket_{MAC}=AP_{MAC}$
            \ElsIf {$PP \text{ is } True$}
                \State $fwdPacket_{MAC}=PP_{MAC}$
            \Else
                \State $fwdPacket_{MAC}=None$
            \EndIf

            \medskip
	\Else
            \medskip
 
            \State $fwdPacket_{MAC}=None$
            \medskip

         \EndIf
            \medskip

         \State \Return $fwdPacket$
	\end{algorithmic}
\end{algorithm}

\subsection{Bounded Delay in networks with an Alternate Parent}\label{bdpc}

\begin{figure}
	\centering
	\includegraphics[scale=0.6]{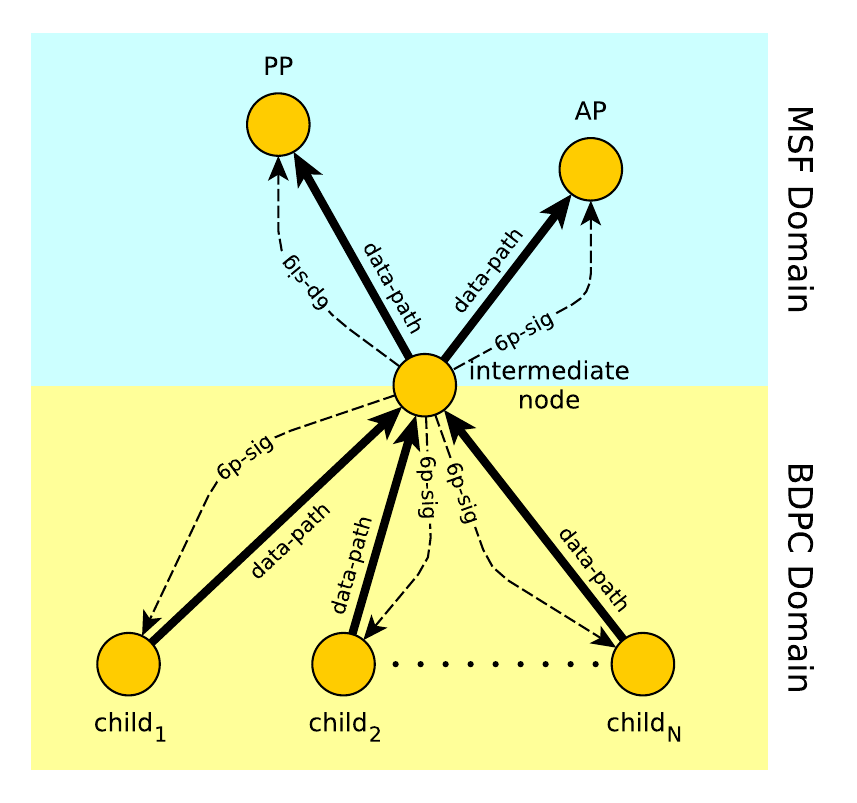}
	\caption{Activity domains of MSF and BDPC. MSF triggers 6P signaling messages to their PP and/or AP because cells need to be added or removed from the schedule due to the outcome of MSF's algorithm (Alg. \ref{alg:msf}). Moreover, BDPC will trigger 6P signalling messages to each of their children because cells need to be added or removed from the schedule due to the outcome of BDPC's algorithm (Alg. \ref{alg:slots}).}
	\label{fig_msf_bdpc}
\end{figure}

According to Aimaretto et al.\cite{AIMARETTO2023100778}, BDPC is agnostic to the topological structure of the network because BDPC manages cells in the parent$\rightarrow$child direction, where each child is identified by its MAC address. On the other hand, if a node manages several parents on the path to the root node to improve the network robustness, the child$\rightarrow$parent resource allocation direction by MSF does not influence the compliance with the application $deadline$. This fact can be observed in Fig. \ref{fig_msf_bdpc}

\begin{figure}
	\centering
	\includegraphics[scale=0.4]{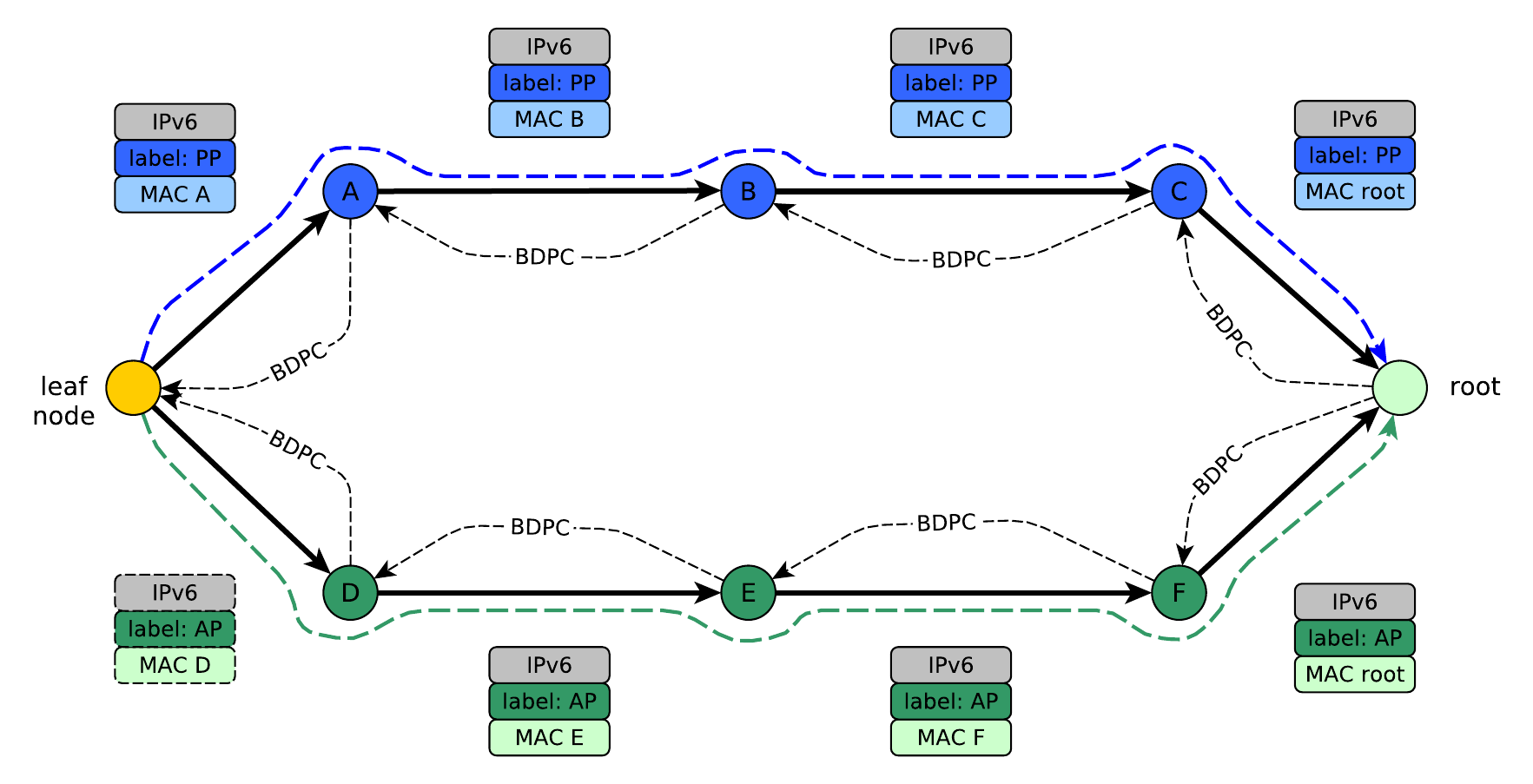}
	\caption{Using MPLS labels, LSPs can be signaled at the source. Since BDPC identifies its' children by their MAC addresses, there will be separate instances of BDPC for each LSP that is built in the network.}
	\label{fig_sr_bdpc}
\end{figure}

Since MPLS labels are used to signal LSPs, the topological tree created by RPL is no longer a single path to the DODAG root, but there are at least two or more paths to the root node: there are as many paths as each intermediate node has alternative parents. Fig. \ref{fig_sr_bdpc} shows an instance of BDPC for each independent LSP.

\section{Simulation Setup}
\label{section:simul_setup}

The 6TiSCH simulator \cite{municio_simulating_2019} includes the standard implementation of the MSF, 6P and RPL protocols in the 6TiSCH stack. The RPL implementation was modified to include the AP selection feature, following the \emph{strict} mode following the description on Section \ref{stateArt}. We also implemented the different \emph{flooding} alternatives, as seen in Section \ref{flooding} and finally, we incorporated BDPC, as seen in Section \ref{bdpc}.

In order to evaluate the performance of our proposal, we executed 6 experiments: 

\begin{itemize}
    \item a standard MSF experiment, which is the reference base;
    \item four experiments, one for each type of \emph{flooding}, as discussed in Section \ref{flooding}. For each case, we used the \emph{strict} implementation for the AP selection;
    \item a final experiment, with the \emph{leafCopy} method and \emph{BDPC}. We used \texttt{sfMax}=10\% and \texttt{sfMin}=5\%. The \emph{strict} implementation is also used for the AP selection.
\end{itemize}

Each experiment was performed with 30 different seeds. The duration for each experiment run was 10000 slotFrames (approximately 2.8 hours in simulation time), to allow the network to stabilize after the initial convergence state.

\begin{figure}[H]
	\centering
	\includegraphics[scale=0.45]{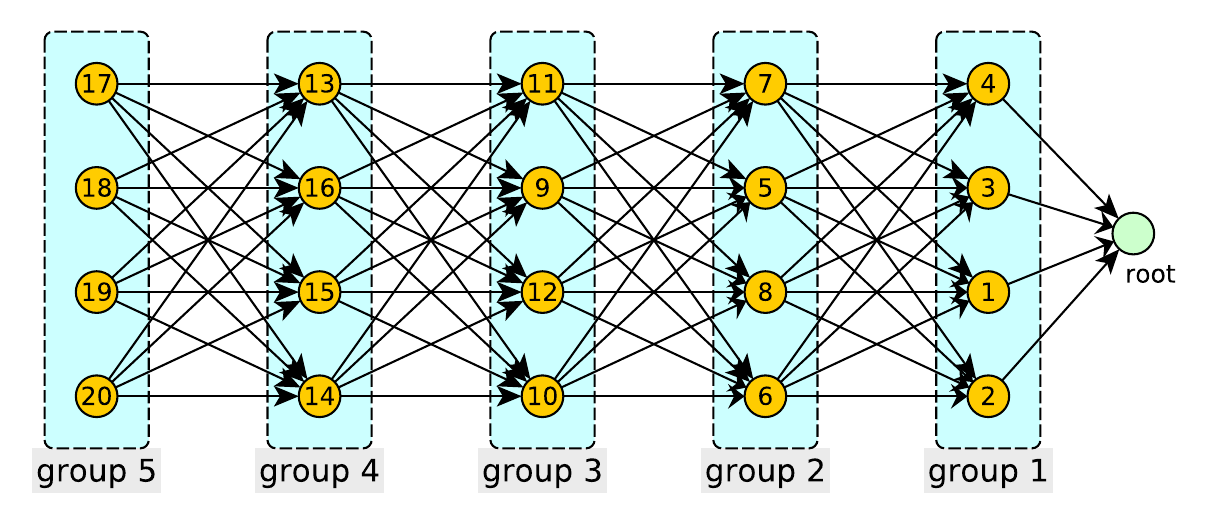}
	\caption{Topology}
	\label{fig:topology}
\end{figure}

The physical topology of the network (Fig. \ref{fig:topology}) is a hierarchy of groups where each group contains four nodes. These groups are set up on the basis of allowing links between nodes. For example, node 9 of group 3 has permitted physical links against nodes 13, 14, 15 and 16 of group 4; and against nodes 5, 6, 7 and 8 of group 2. This means that node 9 can reach both nodes 13 and 6, but node 6 can't reach node 13 and vice-versa: node 13 needs node 9 as a hop to reach node 6. The same logic holds for the rest of the nodes in the network.

In the first group, all nodes have a link to the root node, which can be seen in green. In the rest of the groups, each node has a link to each of the nodes in the groups to the left and right, except for groups 1 and 5. This topology has enough links to neighboring nodes to allow parent change.

Designing the network with this number of fixed hops --given by the different links which form the groups-- allows us to observe the delay that a data packet experiences on its journey from the nodes to the root, especially when packets are generated at the farthest nodes belonging to group 5. The topology can be observed in Fig. \ref{fig:topology}.

The links between nodes are configured with $PDR_{link}$ = 75\% and $RSSI_{link}$ = -91dB, which are representative of imperfect links in IIoT networks in order to appreciate the benefit of having an alternative parent. The relationship between $PDR_{link}$ and $RSSI_{link}$ was obtained experimentally in the model developed by Municio et al. \cite{municio_simulating_2019}. In \cite{jenschke2019alternative, giorgiosLFC}, the authors use values of $PDR_{link}$ of 75\% and 80\%, respectively.

All the nodes (except for the root) generate data traffic where the packet  destination field address points to the root node. This means that there are multiple flows of data packets traversing the network in parallel, from the child nodes towards the root node. Each node generates 90-Byte\footnote{The IEEE 802.15.4 header is 23 bytes long. This leaves room for a 104 bytes payload. A 90-byte payload is an application packet small enough to avoid fragmentation.} data packets that are transmitted every 5s, 10s and 15s\footnote{The default packet transmission period for packets in the simulator is 60s. We've reduced the period to increment the network throughput and test the network under different traffic intensities. In \cite{jenschke2019alternative}, the packet period is fixed to 15s. In \cite{AIMARETTO2023100778} the packet generation period is 30s.}, with 0.05 variance\footnote{The variance of 0.05 is used to  randomly modify the delay of the packet generator as follows: $delay = pkPeridod \times [1+\texttt{uniform}(-pkVar,+pkVar)]$. The rationale behind the variance value choice in the period between packets is that they shall be transmitted within the first available transmission opportunity in the slotFrame and that packet processing time may generate a variable delay on each of the intermediate nodes}. The application maximum delay threshold (\texttt{maxDelay}) is configured to 1.5s\footnote{In control networks, the dead time is the delay a signal takes from a controller output until it is measured and hence there is a response. The effect of dead time in a process needs to be cancelled because time delay makes the designed controller unstable \cite{sunori2017dead}. In \cite{lim1989generalized}, the delay ranges between 30s and 45s. In \cite{leonardi2023mrt} the application deadline is 30s}\footnote{In general, to achieve bounded delay, the duration of the slotFrame shall be shorter or equal to the \texttt{maxDelay} parameter of the application. The minimum timeSlot duration in IEEE 802.15.4 is 10ms \cite{giorgiosLFC}. In case a shorter than 1.5s deadline is part of the requirements, there are two known approaches: either using different length slotFrames, such as in \cite{jenschke2019alternative, chang2016llsf}; or using sub-SlotFrames within a standard slotFrame, such as in \cite{kotsiou2020ldsf}}. The timeSlot duration is 10ms and the slotFrame is composed of 101 timeSlots, both default values from the standard. Table \ref{tab:setup} summarizes the simulation parameters.

\begin{table}[H]
\centering
\caption{Simulation Setup}
\label{tab:setup}
\begin{tabular}{l|r}
\hline
\multicolumn{2}{c}{\textbf{General Parameters}}                                                        \\
\hline
Seeds             & 30 per experiment                                                                  \\
Nodes             & 20 + root                                                                          \\
Topology          & \begin{tabular}[c]{@{}r@{}}Multi-hop with parent\\ change possibility\end{tabular} \\
$PDR_{link}$         & 75.0\%                                                                           \\
$RSSI_{link}$        & -91dB                                                                              \\
Simulation time   & 10000 slotFrames (2.8hs)                                                           \\
slotFrame         & 101 timeSlots                                                                      \\
timeSlot          & 10ms                                                                               \\
Channels          & 16                                                                                 \\
Packet Generation & 5s, 10s, 15s                                                            \\
Packet Variance   & 0.05                                                                \\
Packet Size       & 90 Bytes                                                                           \\
TSCH TX Queue Size     & 10 packets                                                          \\
TSCH Max Retries     & 5                                                                     \\
Application \texttt{maxDelay}     & 1.5s                                                               \\
\hline
\multicolumn{2}{c}{\textbf{MSF standard}}                                                                       \\
\hline
\multicolumn{2}{c}{Original Implementation}     \\
\hline
\multicolumn{2}{c}{\textbf{MSF w/alternate parent support}}           \\
\hline
\multicolumn{2}{c}{strict method}     \\
\hline
\multicolumn{2}{c}{\textbf{BDPC}}                                                                      \\
\hline
\texttt{sfMax}   & 0.1                                                                               \\
\texttt{sfMin}   & 0.05                                                                               \\
PreHop addCell    & 1                                                                                 \\
\hline
\end{tabular}
\end{table}

\section{Results}
\label{section:simul_results}

The results will analyze three metrics and the relationship among them:

\begin{itemize}
    \item $PDR_{e2e}=n_{rx}/n_{tx}$, which measures the packet rate received at the root node to the packets sent by each node as a ratio;
    \item $1-latePaqs_{e2e}=1-n_{delayed}/n_{rx}$, which measures the rate of received packets within the $deadline$, according to Eq. (\ref{latePaqs});
    \item $Time$, which measures the network lifetime, in years. The network lifetime corresponds to the lifetime of the node that depletes first all of the available energy, considering only battery supply.
\end{itemize}

The simulations results are oriented (i) to  compare the benefits of having an alternative parent when using PRE techniques taking into account the lifetime variation according to the \emph{flooding} level in the network; (ii) to observe the effect of BDPC resource management combined with leafCopy to improve $1-latePaqs_{e2e}$. 

\bigskip

\subsection{Enhancing reliability with disjoints paths}\label{sin_bdpc}

\begin{figure}
	\centering
	\includegraphics[scale=0.55]{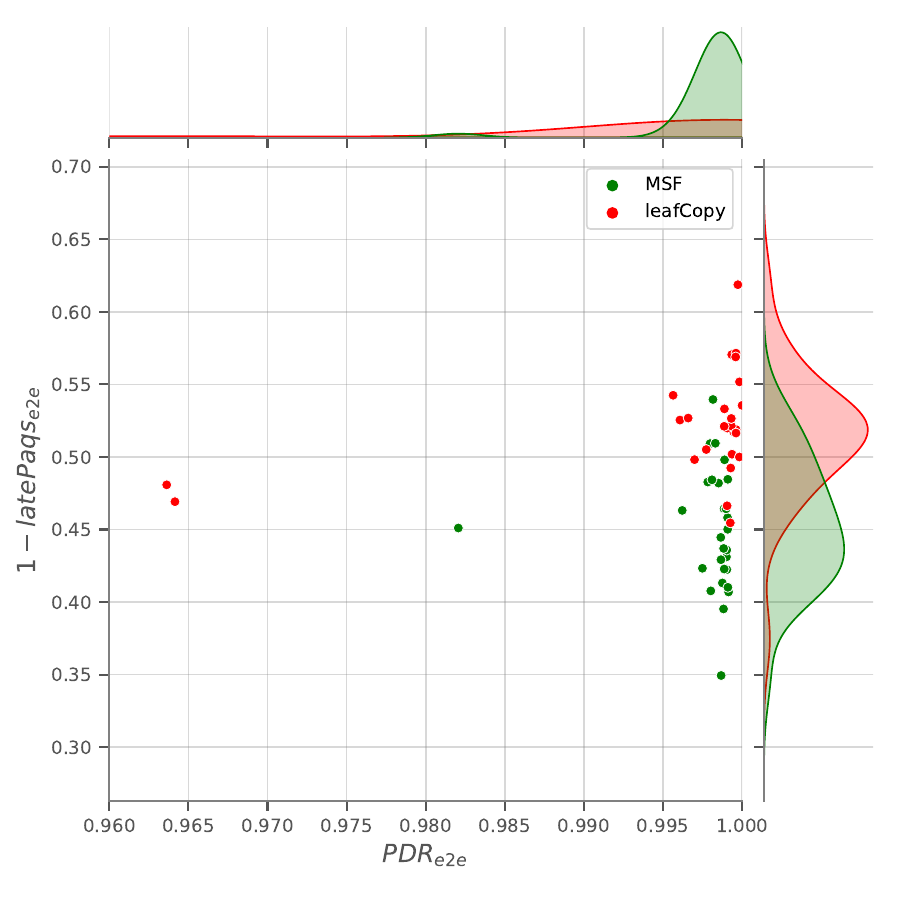}
	\caption{Comparison between MSF and the simplest flooding, LeafCopy. Providing an alternative path to the root improves $1-latePaqs_{e2e}$.}
	\label{fig_msf_leafcopy}
\end{figure}

As a starting point, we analyze the MSF case against the simplest flooding mechanism, \emph{leafCopy}, where packets are only duplicated at the source node. Fig. \ref{fig_msf_leafcopy} shows a joint plot comparing the $PDR_{e2e}$ versus $1-latePaqs_{e2e}$. This figure shows that by providing an alternative path to the root through the alternative parent the result is a deadline-compliant PDR. However, the application requirements are not completely fulfilled yet: while the $PDR_{e2e}$ is higher than 99\% for both cases, there are still between 45\% and 51\% of packets arriving late to the destination, according to the average values displayed in Table \ref{tab_msf_vs_ap_lifetime}.

\begin{figure}
	\centering
	\includegraphics[scale=0.55]{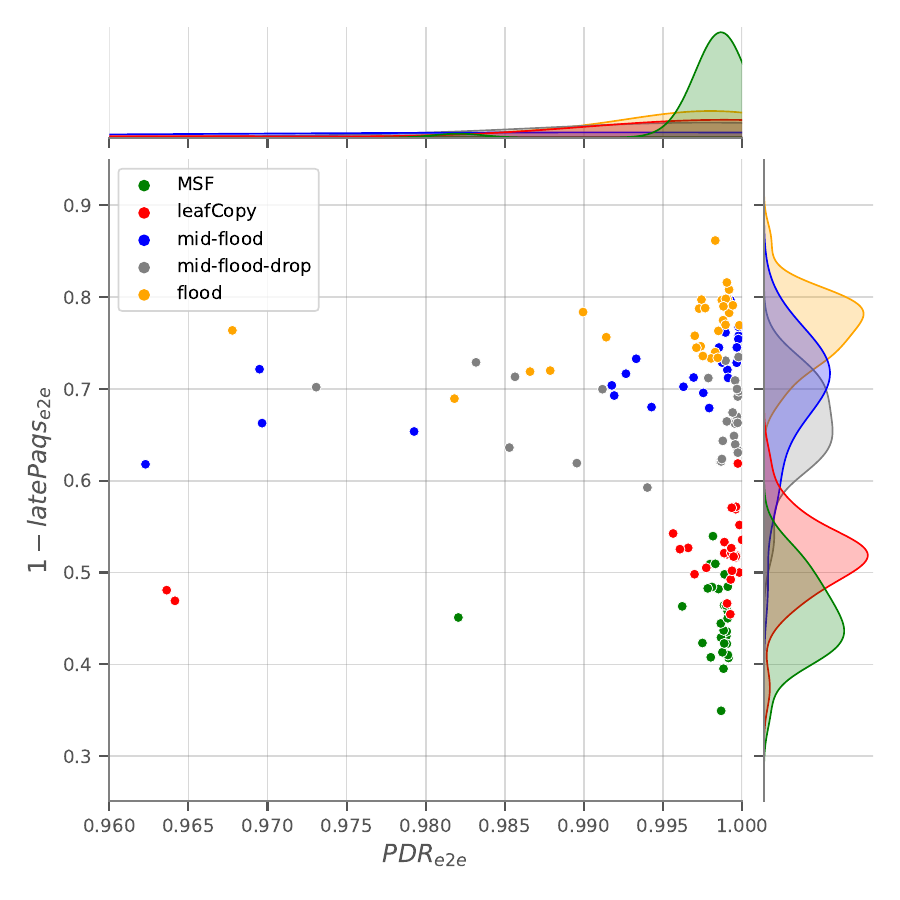}
	\caption{Comparison between MSF and different flooding strategies. As the flooding level increases, the packet delivery on time improves significantly.}
	\label{fig_msf_vs_ap}
\end{figure}

After including packet replication mechanisms such as \emph{mid-flood}, \emph{mid-flood-drop} and \emph{flood}, and compare it to the former analysis, we can observe a better network performance in terms of $1-latePaqs_{e2e}$ metric. Fig. \ref{fig_msf_vs_ap} shows that as packet replication increases, the rate of packets arriving before the $deadline$ increases as well. Thus, for the \emph{flood} case, there are on average 76\% packets arriving before the $deadline$. Table \ref{tab_msf_vs_ap_lifetime} shows the average values in Fig. \ref{fig_msf_vs_ap}.

\begin{figure}
	\centering
	\includegraphics[scale=0.55]{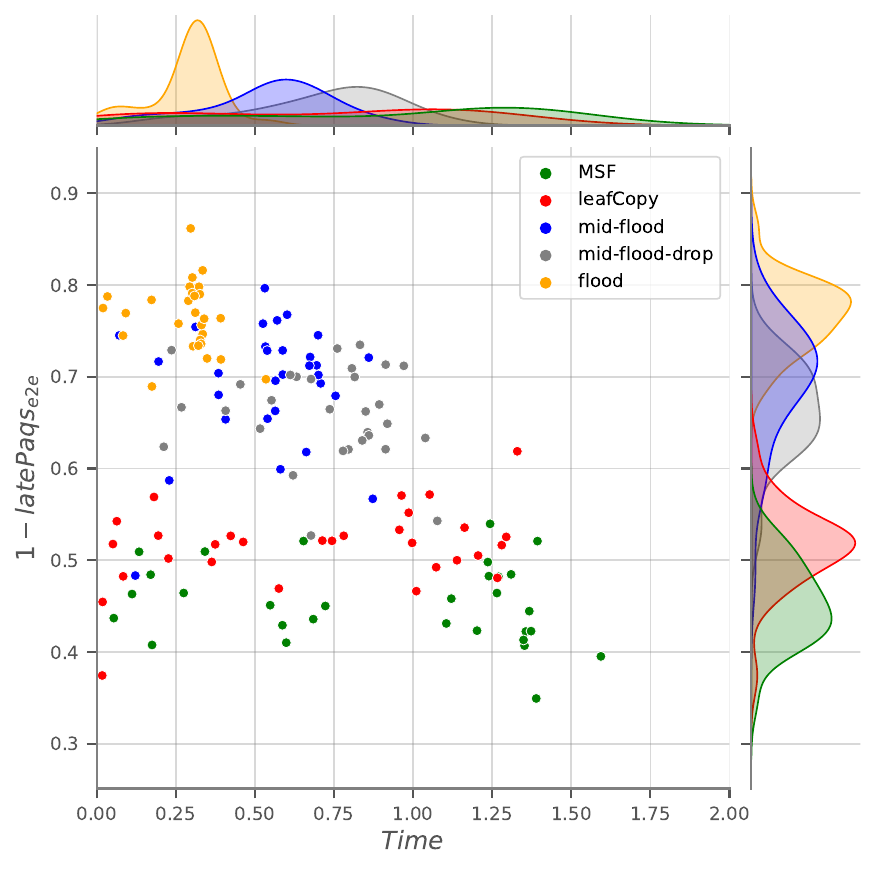}
	\caption{Comparison between MSF and the different flooding strategies, with respect to network lifetime. Average minimum values, in years.}
	\label{fig_msf_vs_ap_lifetime}
\end{figure}

Increasing the flooding level in the network has several consequences. Although the packet replication level offered by the \emph{flood} strategy shows the best performance in terms of packets arriving before the deadline at their destination, it also results in the shortest network lifetime compared to the other alternatives. Fig. \ref{fig_msf_vs_ap_lifetime} shows that both MSF combined with the \emph{leafCopy} and \emph{mid-flood-drop} strategies keep the network operational for the longest time, but offer a worse performance with respect to $1-latePaqs_{e2e}$. Table \ref{tab_msf_vs_ap_lifetime} summarizes the data from the figure and shows the average minimum network lifetime value in years, for each of the strategies.

\begin{table}
\caption{Minimum Average Lifetime (years), $PDR_{e2e}$ and $1-latePaqs_{e2e}$, with respect to the deadline}
\centering
\begin{tabular}{l|ccc}
\hline
Run               & $Time$ & $PDR_{e2e}$ & $1-latePaqs_{e2e}$  \\
\hline
flood          & 0.283740 & 0.993962 & 0.767107       \\
mid-flood      & 0.537890 & 0.972320 & 0.692726       \\
mid-flood-drop & 0.717431 & 0.991169 & 0.659945       \\
leafCopy       & 0.699743 & 0.992505 & 0.515227       \\
MSF                   & 0.907464 & 0.998038 & 0.453725       \\
\hline
\end{tabular}
\label{tab_msf_vs_ap_lifetime}
\end{table}

\begin{figure}
	\centering

\begin{subfigure}{.49\linewidth}
\includegraphics[width=\textwidth]{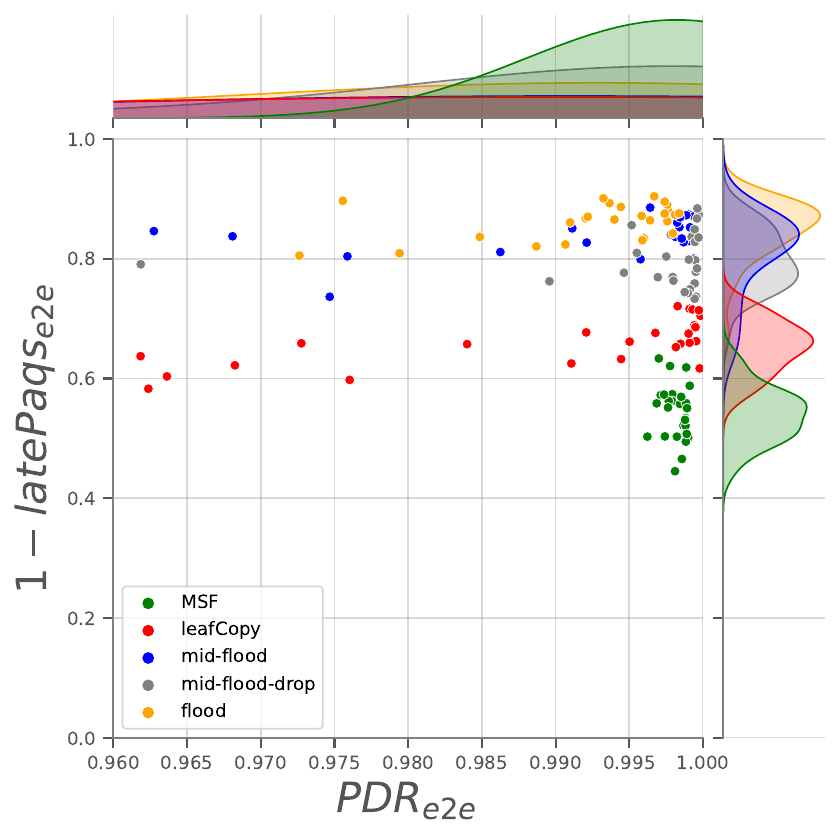}%
\caption{Packet/5s}%
\label{fig_joint_pk05_sinBDPC}%
\end{subfigure}
\hfill%
\begin{subfigure}{.49\linewidth}
\includegraphics[width=\textwidth]{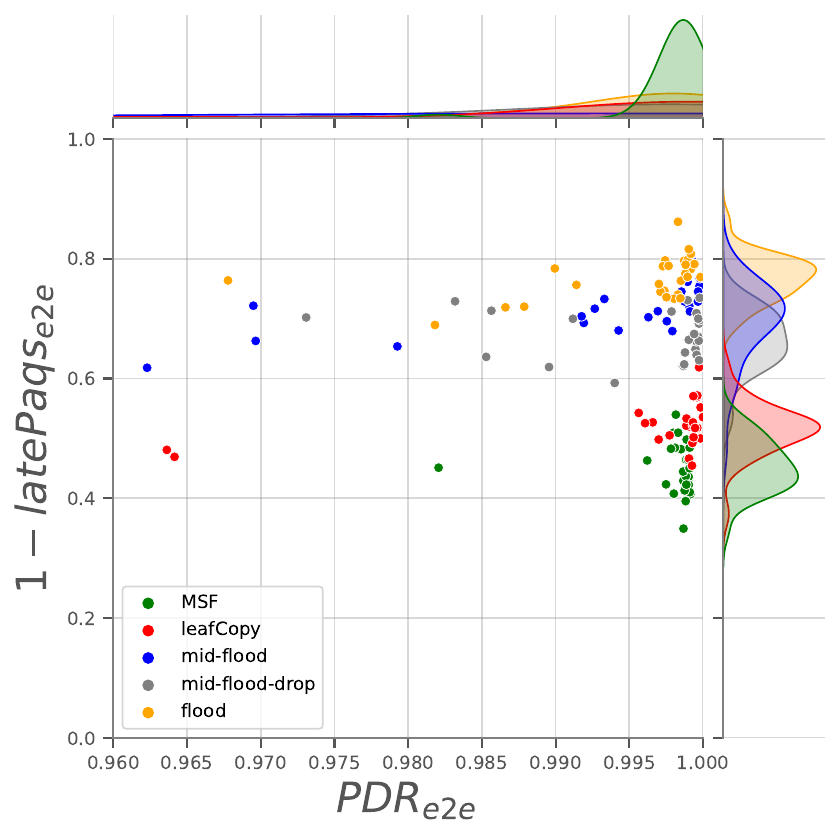}%
\caption{Packet/10s}%
\label{fig_joint_pk10_sinBDPC}%
\end{subfigure}
\hfill%

\bigskip
\begin{subfigure}{.49\linewidth}
\includegraphics[width=\textwidth]{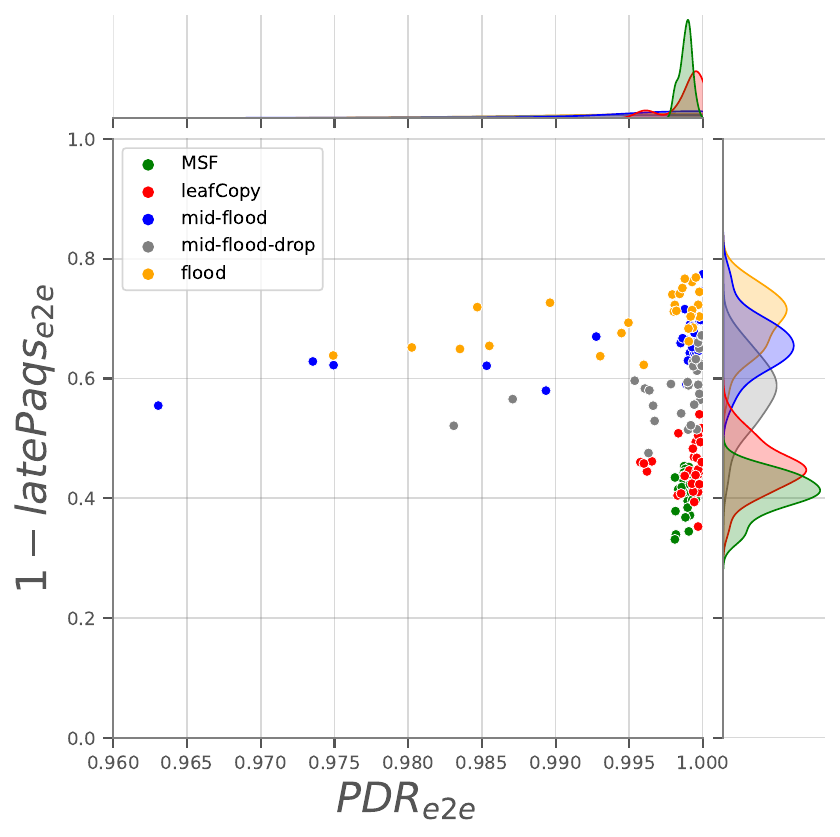}%
\caption{Packet/15s}%
\label{fig_joint_pk15_sinBDPC}%
\end{subfigure}%

	\caption{Comparison between standard MSF and different flooding strategies, considering different packet generation periods.}

    \label{fig_msf_vs_ap_allPk}
\end{figure}

Fig. \ref{fig_msf_vs_ap_allPk} shows the network performance considering the full range of packet generation periods. In general, for a small packet generation period, i.e., a higher throughput, the deadline-compliant PDR is higher than larger periods. This is a consequence to the predefined resource reservation frequency in the slotFrame: the higher the throughput, the higher the packet reception per unit of time and therefore the lower the reaction of Alg. \ref{alg:msf} from MSF, avoiding changes on the schedule.


\bigskip

\subsection{Enhancing determinism with BDPC over disjoint paths}

In Section \ref{sin_bdpc}, we described how $1-latePaqs_{e2e}$ improves when we enable the use of alternate parents. We also observed that $1-latePaqs_{e2e}$ further improves with an increase of packet copies, albeit reducing the network lifetime. Conversely, in order to increase the network lifetime and improve its robustness, the packet flooding scheme must be exchanged with a less aggressive one: the \emph{leafCopy} scheme results into the extension of network lifetime and improves robustness, creating disjoint paths to the root. But \emph{leafCopy} has a only slightly better (13\%) performance in terms of $1-latePaqs_{e2e}$ when compared to MSF, since \emph{leafCopy} is equivalent to the use of two independent instances of MSF in parallel, one for each LSP.

In order to solve the \emph{leafCopy} behavior described above, we propose to combine BDPC with the \emph{leafCopy} mechanism to generate parallel paths signaled at the leafNode, and aggregate resources based on the application deadline using BDPC (Fig. \ref{fig_sr_bdpc}).

\begin{figure}
	\centering
	\includegraphics[scale=0.55]{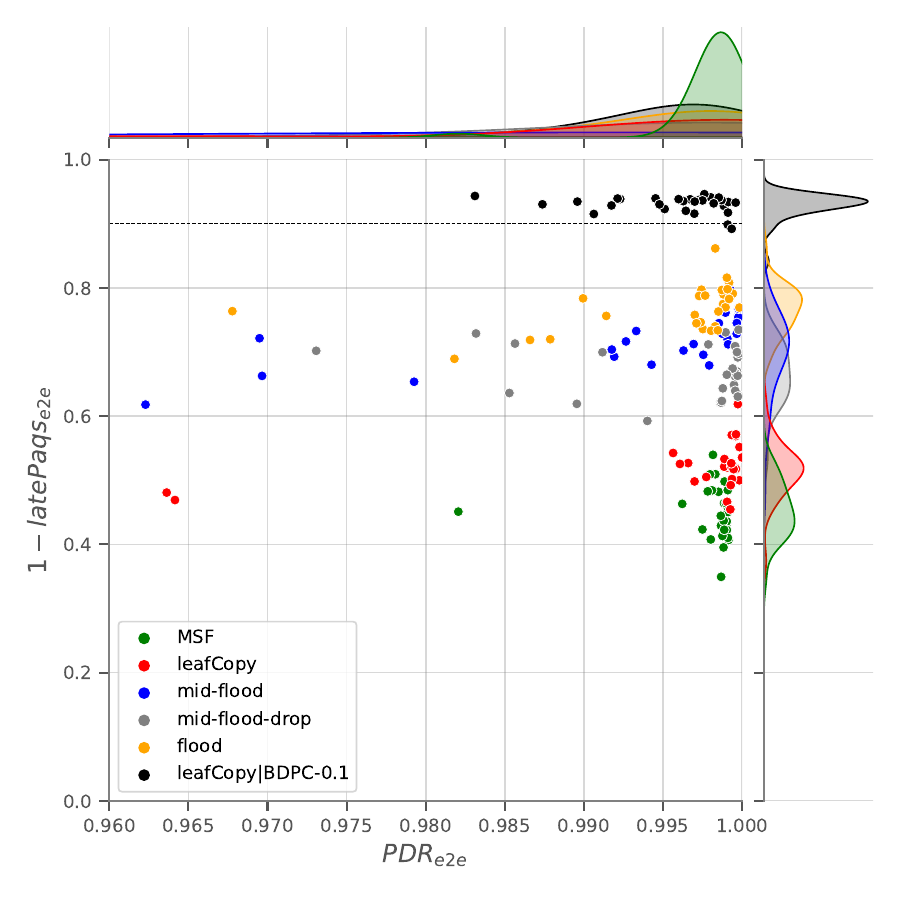}
	\caption{BDPC in conjunction with the \emph{leafCopy} strategy. The dashed horizontal line represents the value $1-\texttt{sfMax}$: this is the rate of packets arriving within the deadline, when BDPC is used. }
	\label{fig_msf_ap_bdpc}
\end{figure}

Fig. \ref{fig_msf_ap_bdpc} shows the network performance when using BDPC for a \texttt{sfMax}=0.1 value\footnote{The value \texttt{sfMax}=0.1 guarantees that at least 90\% of the packets will arrive before the deadline because, according to Eq. (\ref{bdpc_prob}), the packet arrival probability before the application deadline is 1-\texttt{sfMax}}. We also highlight that adding the BDPC variant to the \emph{leafCopy} approach generates, first, a parallel path to improve the robustness of the network (Fig. \ref{fig_sr_bdpc}) and, in addition, improves the PDR within the application deadline by 2.04 times with respect to the reference, MSF. Moreover, the leafCopy+BDPC combination is superior to the \emph{flood} scheme since it delivers 21\% more packets within the $deadline$ and the network lives --on average-- almost 1.5 times longer. Table \ref{tab_msf_vs_ap_lifetime_bdpc} shows the average values of $PDR_{e2e}$ versus $1-latePaqs_{e2e}$. 


\begin{figure}
	\centering
	\includegraphics[scale=0.55]{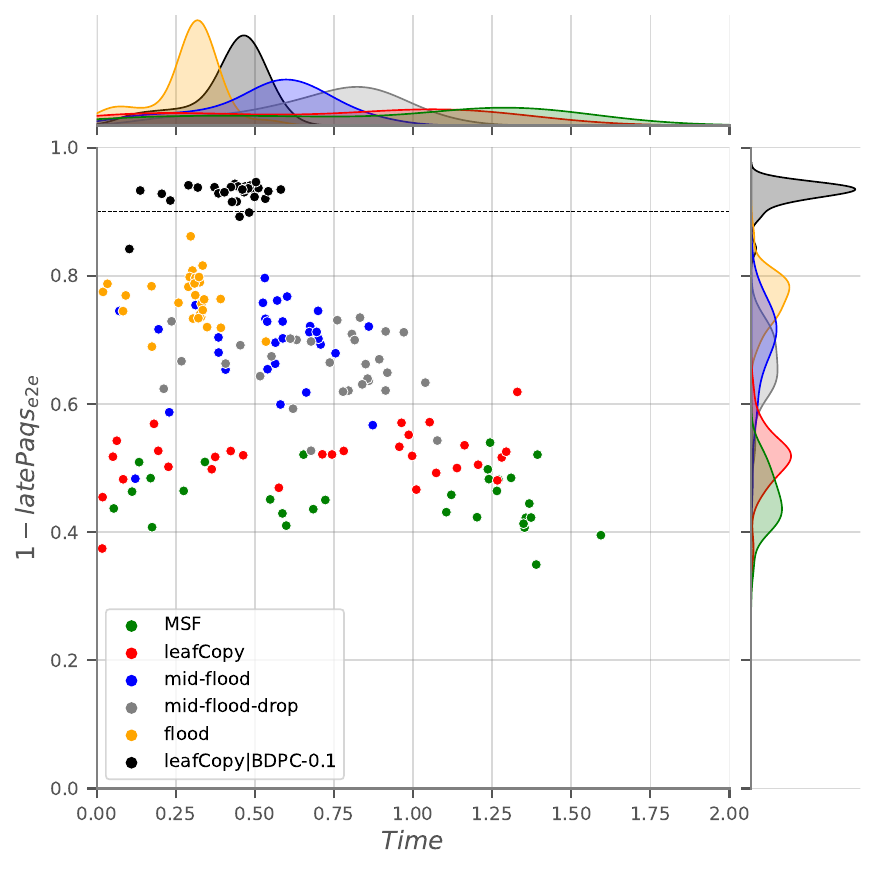}
	\caption{When BDPC is enabled, the resource allocation strategy meets the realtime application requirements. By adding resources, the network lifetime is shorter due to the increase in cell usage. However, since resource allocation is governed by Alg. \ref{alg:slots}, the network lifetime will be longer with respect to an uncontrolled PRE behavior condition.}
	\label{fig_msf_ap_bdpc_lifetime}
\end{figure}

Fig. \ref{fig_msf_ap_bdpc_lifetime} depicts the network lifetime behavior: On the one hand, even though MSF or leafCopy-original provide independently a longer lifetime, they do not comply with the application deadline requirements. On the other hand, BDPC complies with the application deadline requirements at the expense of a shorter network lifetime. PRE algorithms such as flood or mid-flood improve ($1-latePaqs_{e2e}$), but the network lifetime is shorter than the use of BDPC alone. Finally, the combination of BDPC and leafCopy shows the best balance between compliance with application deadline and network lifetime.



To summarize, Table \ref{tab_final_comparison} provides a comparison of the combined BDPC+leafCopy solution versus: (i) the standard reference, MSF; (ii) the case using \emph{leafCopy}; (iii) the case using \emph{flood}.

The combined BDPC+leafCopy solution is superior in terms of $1-latePaqs_{e2e}$ for all cases, and despite the fact that the network lifetime is shorter than MSF or the original leafCopy, the latter cannot guarantee on-time delivery of critical packets depending on the application deadline. Finally, when the solution is compared against \emph{flood}, the BDPC+leafCopy solution delivers 21\% more packets and lives almost 50\% longer.

\begin{figure}
	\centering

\begin{subfigure}{.49\linewidth}
\includegraphics[width=\columnwidth]{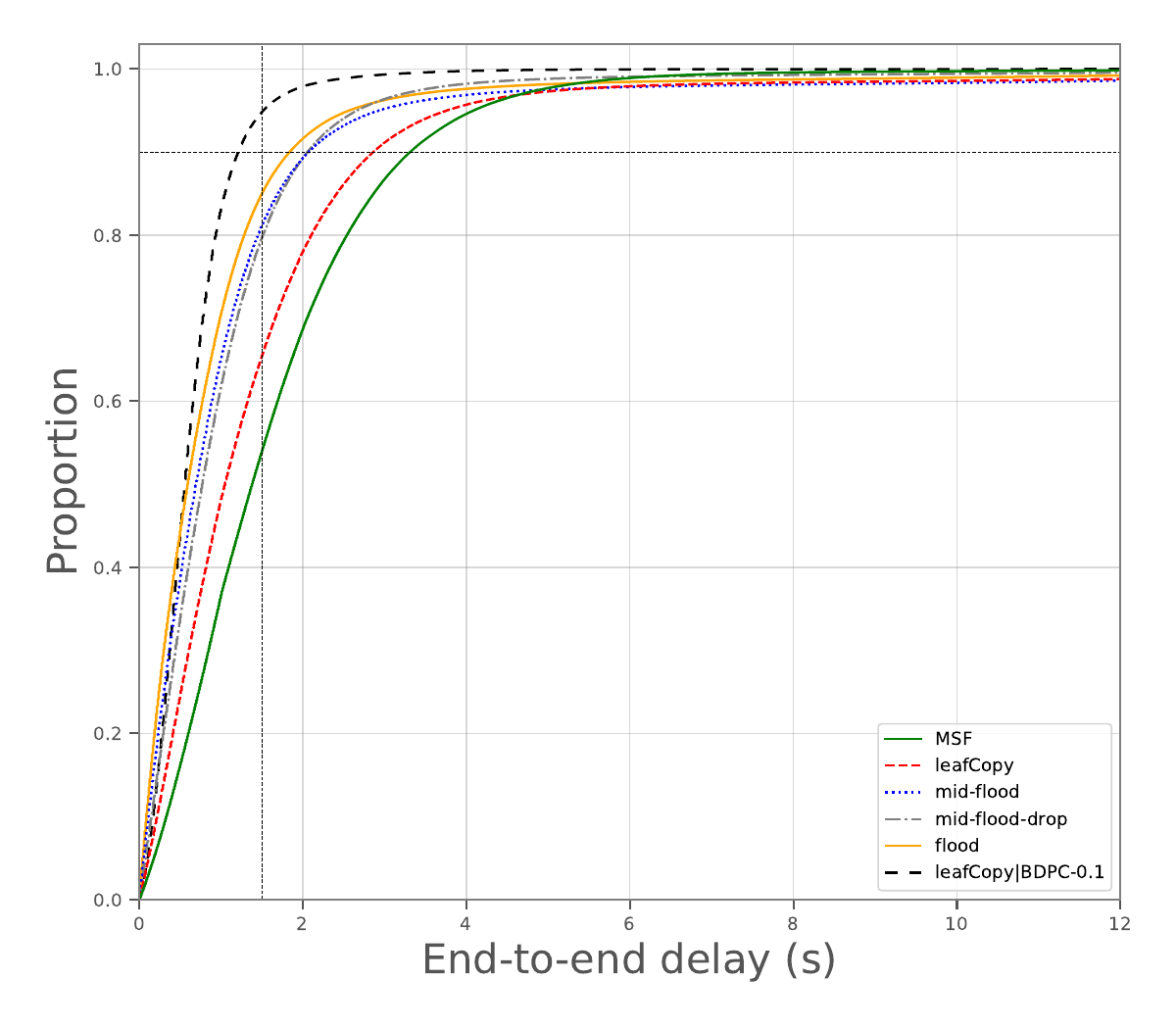}%
\caption{Packet/5s}%
\label{fig_ecdf_pk05}%
\end{subfigure}
\hfill%
\begin{subfigure}{.49\linewidth}
\includegraphics[width=\columnwidth]{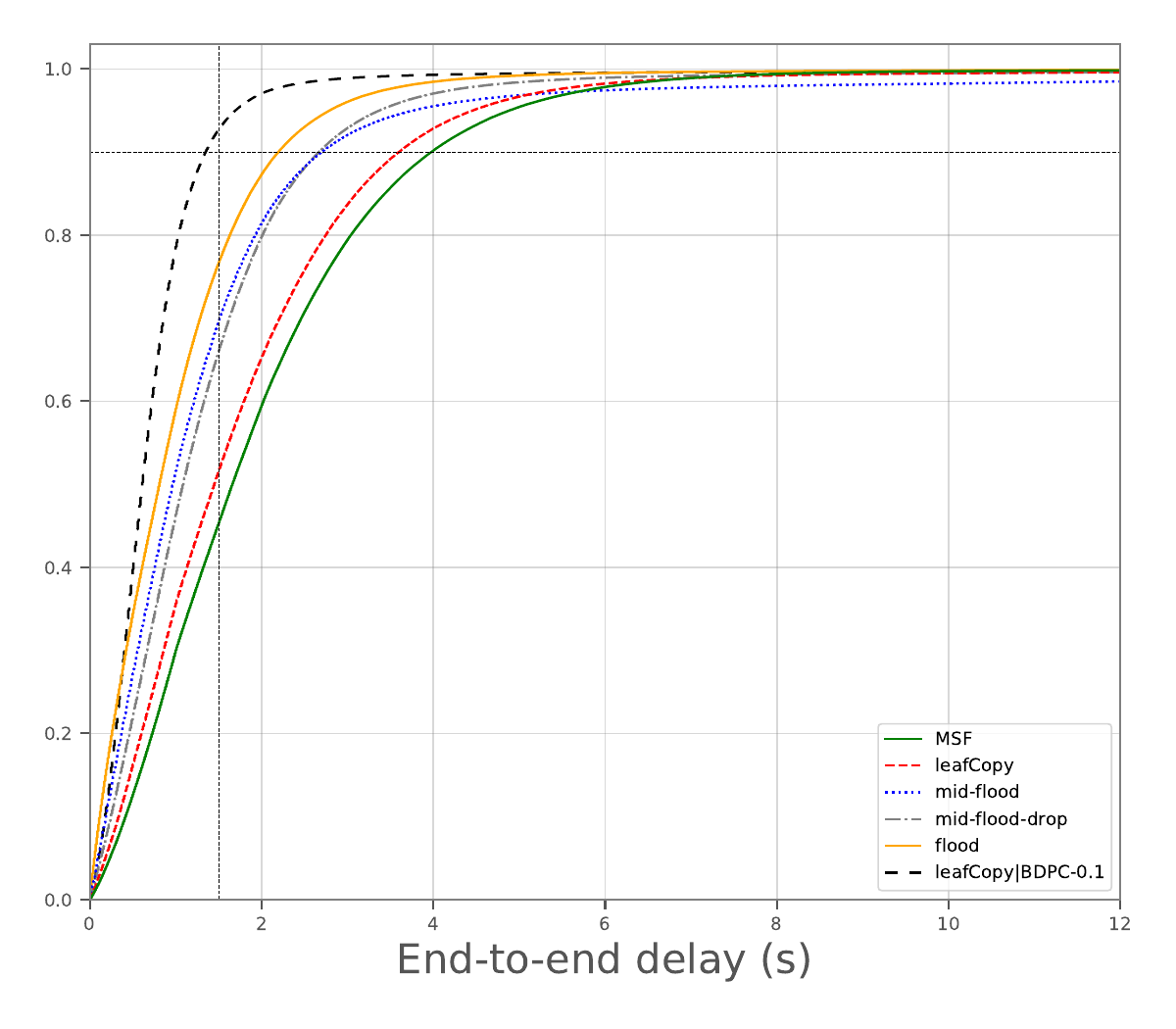}%
\caption{Packet/10s}%
\label{fig_ecdf_pk10}%
\end{subfigure}
\hfill%

\bigskip
\begin{subfigure}{.49\linewidth}
\includegraphics[width=\columnwidth]{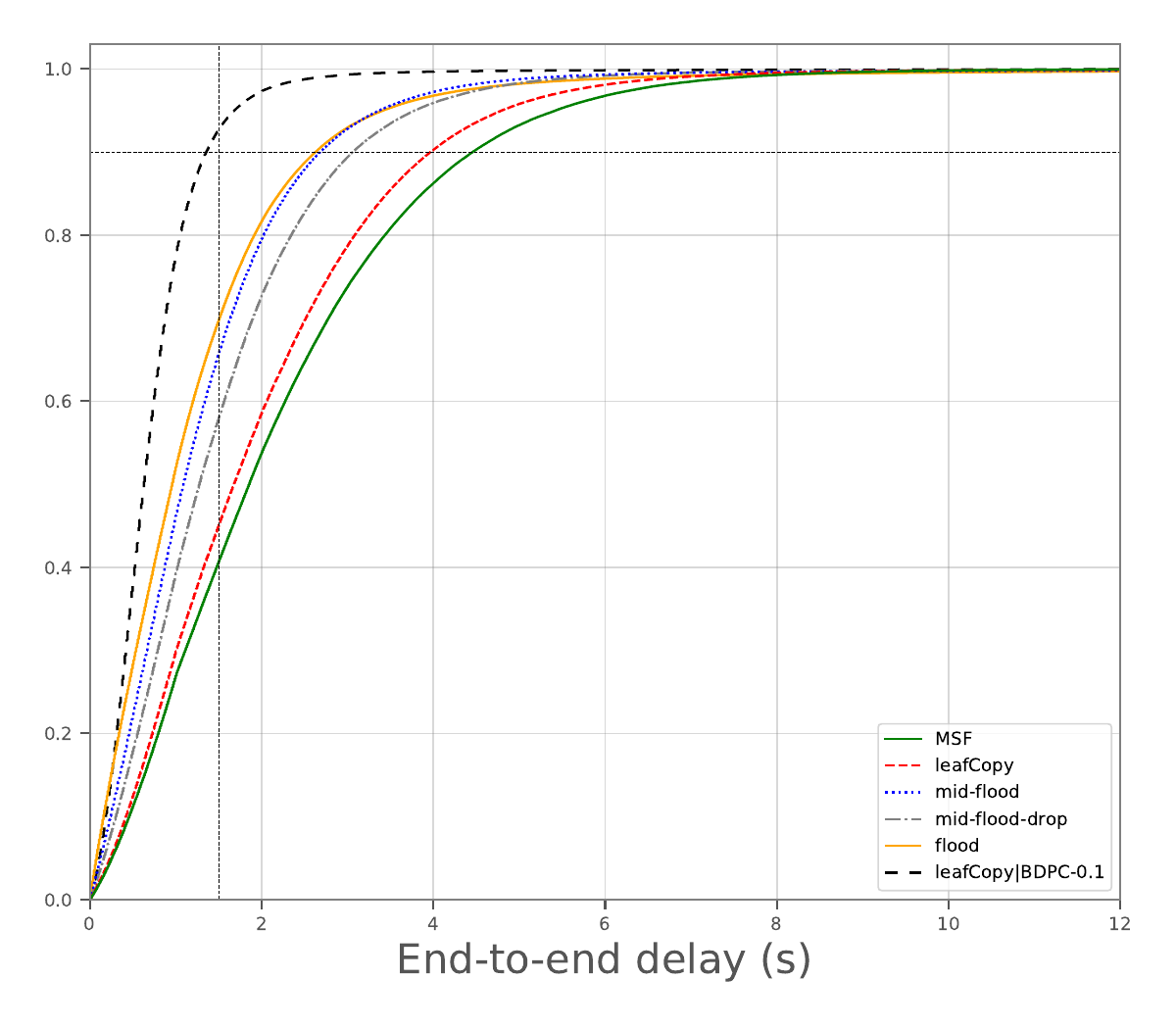}%
\caption{Packet/15s}%
\label{fig_ecdf_pk15}%
\end{subfigure}%

	\caption{End-to-End data packets delay. The vertical black-dashed line represents the maximum application delay, \texttt{maxDelay}. The horizontal dashed line represents the value of $1-\texttt{sfMax}$, for \texttt{sfMax} 10\%. According to Eq. (\ref{bdpc_prob}), BDPC guarantees that the PDR before de the application deadline will be $1-\texttt{sfMax}$ irrespective of the source traffic pattern.}
	\label{fig_e2e_delay}
\end{figure}

Fig. \ref{fig_e2e_delay} shows the cumulative end-to-end delay distribution for each of the experiments considering the three traffic intensities. The vertical black dashed line represents the maximum delay tolerated by the application. The horizontal black dashed line represents the $1-\texttt{sfMax}$ value. The delay is measured once the packets are received at the root and calculated as the difference between the packet reception time at the root minus the transmission time at the leaf node, $d=t_{rx}-t_{tx}$. In this case, we can observe that with the use of BDPC, the packets arriving at the root node are always guaranteed to comply with the maximum allowed delay, depending on the value \texttt{sfMax}.

\begin{figure}
	\centering

\begin{subfigure}{.49\linewidth}
\includegraphics[width=\columnwidth]{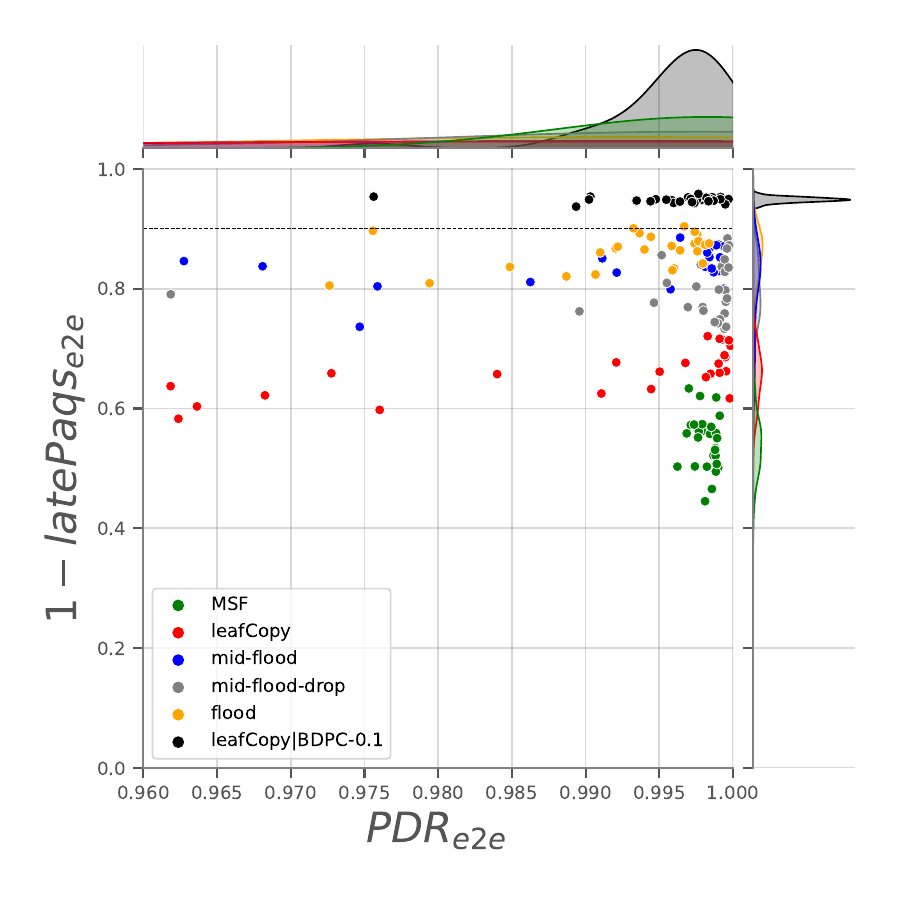}%
\caption{Packet/5s}%
\label{fig_joint_pk05}%
\end{subfigure}
\hfill%
\begin{subfigure}{.49\linewidth}
\includegraphics[width=\columnwidth]{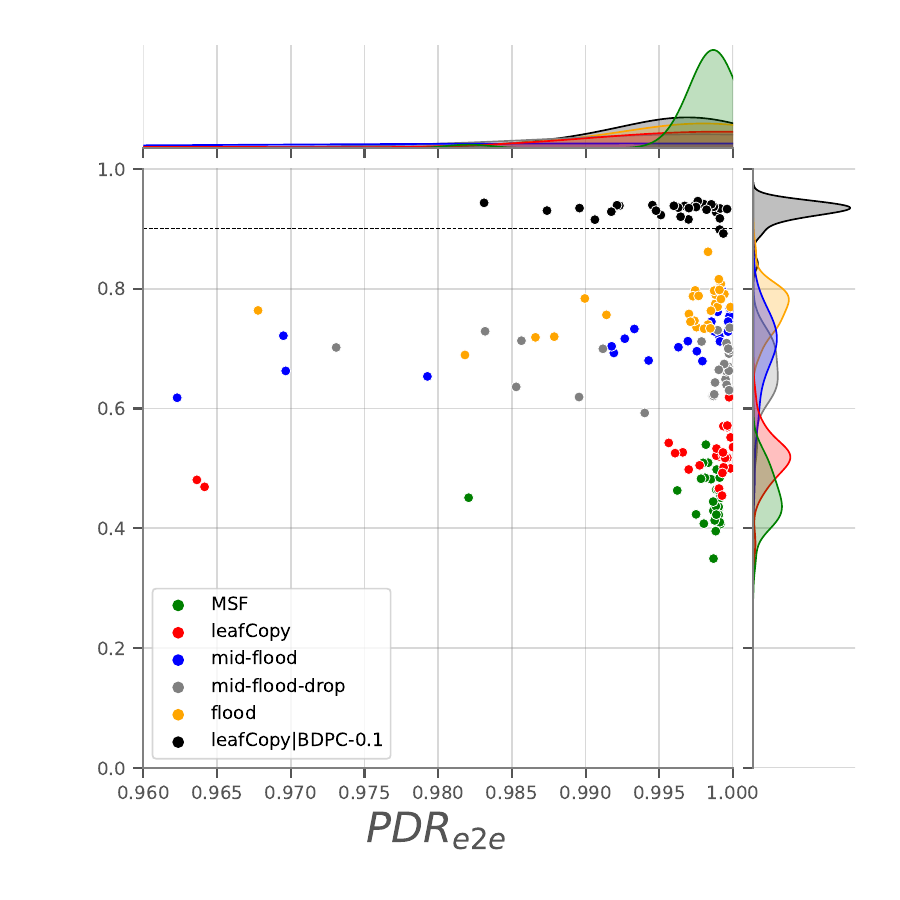}%
\caption{Packet/10s}%
\label{fig_joint_pk10}%
\end{subfigure}
\hfill%

\bigskip
\begin{subfigure}{.49\linewidth}
\includegraphics[width=\columnwidth]{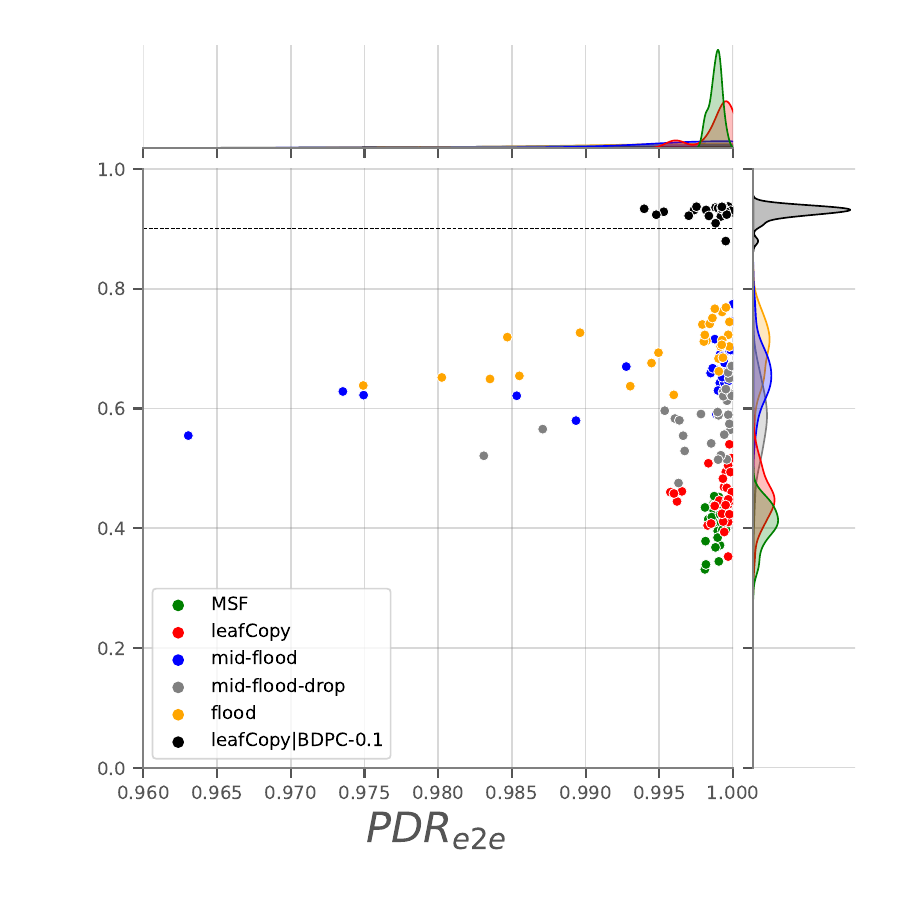}%
\caption{Packet/15s}%
\label{fig_joint_pk15}%
\end{subfigure}%

	\caption{LeafCopy improves reliability due to the availability of alternate paths towards the root node. However, BDPC guarantees that the rate of packets arriving before the application deadline at the root node will be above $1-\texttt{sfMax}$ irrespective of the traffic pattern. The combination of LeafCopy with BDPC improves both reliability and determinism simultaneously.}
	\label{fig_msf_vs_ap_allPk_bdpc}
\end{figure}

Finally, Fig. \ref{fig_msf_vs_ap_allPk_bdpc} shows the performance of $1-latePaqs_{e2e}$ for different traffic intensities. The use of BDPC guarantees the delivery of packets before the application deadline, as defined by the parameter \texttt{sfMax}, regardless of the traffic pattern. This is possible because BDPC is agnostic to the type of traffic pattern, since BDPC reacts in a per-packet basis. For the rest of the cases, although the distribution of delivery time improves by increasing the flooding level, such delivery stategy causes an increase in energy consumption which grows with the size of the network.

\begin{table}
\centering
\caption{Minimum Average Lifetime (years), $PDR_{e2e}$ and $1-latePaqs_{e2e}$, when Deadline is considered and BDPC is used together with leafCopy.}
\label{tab_msf_vs_ap_lifetime_bdpc}
\begin{tabular}{l|ccc}
\hline
Run               & $Time$ & $PDR_{e2e}$ & $1-latePaqs_{e2e}$  \\
\hline
leafCopy+BDPC & 0.416986 & 0.994282 & 0.927396 \\
flood             & 0.283740 & 0.993962 & 0.767107 \\
mid-flood         & 0.537890 & 0.972320 & 0.692726 \\
mid-flood-drop    & 0.717431 & 0.991169 & 0.659945 \\
leafCopy          & 0.699743 & 0.992505 & 0.515227 \\
MSF               & 0.907464 & 0.998038 & 0.453725 \\
\hline
\end{tabular}
\end{table}

\begin{table}
\centering
\caption{Average relative comparison of leafCopy+BDPC against MSF, leafCopy and flood.}
\begin{tabular}{l|cc}
\hline
Run      & $Time$   & $1-latePaqs_{e2e}$ \\
\hline
leafCopy+BDPC vs flood    & 1.4696 & 1.2090     \\
leafCopy+BDPC vs leafCopy & 0.5959 & 1.8000     \\
leafCopy+BDPC vs MSF      & 0.4595 & 2.0440     \\
\hline
\end{tabular}

\label{tab_final_comparison}
\end{table}

\section{Conclusion}\label{section:conclusion}

This paper proposes a solution combining link and routing layer strategies to improve network robustness and network determinism and providing a deadline-compliant PDR at the root node, according to the requirements of industrial IIoT applications.

In IIoT, at least three realtime flow requirements must be met: (i) a very high packet delivery rate at the destination, or $PDR_{e2e}$; (ii) a very high packet delivery rate before the maximum deadline defined by the application, or $1-latePaqs_{e2e}$; (iii) the longest network lifetime possible, especially if the nodes are battery-powered. 

Our proposal provides a new solution to IIoT requirements by first, enabling the use of an alternate parent in a distributed resource allocation network; second, by reducing power consumption using disjoint paths for data flows based on MPLS mechanisms at the leaf-node and third, by leveraging BDPC resource allocation to overcome link and node variations in packet flow capacity and deliver critical packets to the destination before a maximum application deadline. 

Consequently, by combining this set of strategies, the IIoT network can deliver 2.04 times more packets before the application deadline compared to the current standard implementation and can save up to 50\% energy compared to a pure PRE-based strategy, where packet replication is performed hop-by-hop, without any control.

\section*{Acknowledgment}
The authors would like to thank the CYTED AgIoT Project (520rt0011), CORFO CoTH2O ‘‘Consorcio de Gestión de Recursos Hídricos para la Macrozona Centro-Sur’’ (20CTECGH-145896), Proyecto Asociativo UDP ‘‘Plataformas Digitales como modelo organizacional’’ for their support during the development of this research work.


 \bibliographystyle{elsarticle-num} 
 \bibliography{cas-refs}

\begin{thebibliography}{10}
\expandafter\ifx\csname url\endcsname\relax
  \def\url#1{\texttt{#1}}\fi
\expandafter\ifx\csname urlprefix\endcsname\relax\def\urlprefix{URL }\fi
\expandafter\ifx\csname href\endcsname\relax
  \def\href#1#2{#2} \def\path#1{#1}\fi

\bibitem{thubert2017converging}
P.~Thubert, Converging over deterministic networks for an industrial internet,
  Ph.D. thesis, Ecole nationale sup{\'e}rieure Mines-T{\'e}l{\'e}com Atlantique
  (2017).

\bibitem{koutsiamanis2018best}
R.-A. Koutsiamanis, G.~Z. Papadopoulos, X.~Fafoutis, J.~M. Del~Fiore,
  P.~Thubert, N.~Montavont, From best effort to deterministic packet delivery
  for wireless industrial iot networks, IEEE Transactions on Industrial
  Informatics 14~(10) (2018) 4468--4480.

\bibitem{draft-ietf-raw-use-cases}
C.~J. Bernardos, G.~Z. Papadopoulos, P.~Thubert, F.~Theoleyre,
  \href{https://datatracker.ietf.org/doc/draft-ietf-raw-use-cases/11/}{{RAW
  Use-Cases}}, Internet-Draft draft-ietf-raw-use-cases-11, Internet Engineering
  Task Force, work in Progress (Apr. 2023).
\newline\urlprefix\url{https://datatracker.ietf.org/doc/draft-ietf-raw-use-cases/11/}

\bibitem{jenschke2020toward}
T.~L. Jenschke, Toward reliable and bounded latency for internet of things,
  Ph.D. thesis, Ecole nationale sup{\'e}rieure Mines-T{\'e}l{\'e}com Atlantique
  (2020).

\bibitem{AIMARETTO2023100778}
L.~Aimaretto, D.~Dujovne,
  \href{https://www.sciencedirect.com/science/article/pii/S2542660523001014}{Bdpc:
  Controlling application delay in 6tisch networks for the industrial internet
  of things}, Internet of Things (2023) 100778\href
  {https://doi.org/https://doi.org/10.1016/j.iot.2023.100778}
  {\path{doi:https://doi.org/10.1016/j.iot.2023.100778}}.
\newline\urlprefix\url{https://www.sciencedirect.com/science/article/pii/S2542660523001014}

\bibitem{ieee2015ieee}
IEEE, Ieee 802.15. 4-2015-ieee standard for low-rate wireless networks, IEEE
  standards association (2015).

\bibitem{urke2021survey}
A.~R. Urke, {\O}.~Kure, K.~{\O}vsthus, A survey of 802.15. 4 tsch schedulers
  for a standardized industrial internet of things, Sensors 22~(1) (2021) 15.

\bibitem{draft-ietf-raw-architecture}
P.~Thubert, G.~Z. Papadopoulos,
  \href{https://datatracker.ietf.org/doc/html/draft-ietf-raw-architecture-04}{{Reliable
  and Available Wireless Architecture}}, Internet-Draft
  draft-ietf-raw-architecture-04, Internet Engineering Task Force, work in
  Progress (Mar. 2022).
\newline\urlprefix\url{https://datatracker.ietf.org/doc/html/draft-ietf-raw-architecture-04}

\bibitem{rfc9030}
P.~Thubert, \href{https://www.rfc-editor.org/info/rfc9030}{{An Architecture for
  IPv6 over the Time-Slotted Channel Hopping Mode of IEEE 802.15.4 (6TiSCH)}},
  RFC 9030 (May 2021).
\newblock \href {https://doi.org/10.17487/RFC9030}
  {\path{doi:10.17487/RFC9030}}.
\newline\urlprefix\url{https://www.rfc-editor.org/info/rfc9030}

\bibitem{rfc9033}
T.~Chang, M.~Vučinić, X.~Vilajosana, S.~Duquennoy, D.~R. Dujovne,
  \href{https://www.rfc-editor.org/info/rfc9033}{{6TiSCH Minimal Scheduling
  Function (MSF)}}, RFC 9033 (May 2021).
\newblock \href {https://doi.org/10.17487/RFC9033}
  {\path{doi:10.17487/RFC9033}}.
\newline\urlprefix\url{https://www.rfc-editor.org/info/rfc9033}

\bibitem{rfc8480}
Q.~Wang, X.~Vilajosana, T.~Watteyne,
  \href{https://www.rfc-editor.org/info/rfc8480}{{6TiSCH Operation Sublayer
  (6top) Protocol (6P)}}, RFC 8480 (Nov. 2018).
\newblock \href {https://doi.org/10.17487/RFC8480}
  {\path{doi:10.17487/RFC8480}}.
\newline\urlprefix\url{https://www.rfc-editor.org/info/rfc8480}

\bibitem{rfc6550}
R.~Alexander, A.~Brandt, J.~Vasseur, J.~Hui, K.~Pister, P.~Thubert, P.~Levis,
  R.~Struik, R.~Kelsey, T.~Winter,
  \href{https://www.rfc-editor.org/info/rfc6550}{{RPL: IPv6 Routing Protocol
  for Low-Power and Lossy Networks}}, RFC 6550 (Mar. 2012).
\newblock \href {https://doi.org/10.17487/RFC6550}
  {\path{doi:10.17487/RFC6550}}.
\newline\urlprefix\url{https://www.rfc-editor.org/info/rfc6550}

\bibitem{ryu2005urgency}
S.~Ryu, B.~Ryu, H.~Seo, M.~Shin, Urgency and efficiency based packet scheduling
  algorithm for ofdma wireless system, in: IEEE International Conference on
  Communications, 2005. ICC 2005. 2005, Vol.~4, IEEE, 2005, pp. 2779--2785.

\bibitem{jenschke2019alternative}
T.~L. Jenschke, G.~Z. Papadopoulos, R.-A. Koutsiamanis, N.~Montavont,
  Alternative parent selection for multi-path rpl networks, in: 2019 IEEE 5th
  world forum on internet of things (WF-IoT), IEEE, 2019, pp. 533--538.

\bibitem{draft-ietf-roll-nsa-extension-10}
R.-A. Koutsiamanis, G.~Z. Papadopoulos, N.~Montavont, P.~Thubert,
  \href{https://datatracker.ietf.org/doc/html/draft-ietf-roll-nsa-extension-10}{{Common
  Ancestor Objective Function and Parent Set DAG Metric Container Extension}},
  Internet-Draft draft-ietf-roll-nsa-extension-10, Internet Engineering Task
  Force, work in Progress (Oct. 2020).
\newline\urlprefix\url{https://datatracker.ietf.org/doc/html/draft-ietf-roll-nsa-extension-10}

\bibitem{giorgiosPacketDuplicationEnergy}
G.~Z. Papadopoulos, J.~Beaudaux, A.~Gallais, P.~Chatzimisios, T.~Noël, Toward
  a packet duplication control for opportunistic routing in wsns, in: 2014 IEEE
  Global Communications Conference, 2014, pp. 94--99.
\newblock \href {https://doi.org/10.1109/GLOCOM.2014.7036790}
  {\path{doi:10.1109/GLOCOM.2014.7036790}}.

\bibitem{giorgiosLFC}
R.-A. Koutsiamanis, G.~Z. Papadopoulos, X.~Fafoutis, J.~M.~D. Fiore,
  P.~Thubert, N.~Montavont, From best effort to deterministic packet delivery
  for wireless industrial iot networks, IEEE Transactions on Industrial
  Informatics 14~(10) (2018) 4468--4480.
\newblock \href {https://doi.org/10.1109/TII.2018.2856884}
  {\path{doi:10.1109/TII.2018.2856884}}.

\bibitem{rfc3031}
A.~Viswanathan, E.~C. Rosen, R.~Callon,
  \href{https://www.rfc-editor.org/info/rfc3031}{{Multiprotocol Label Switching
  Architecture}}, RFC 3031 (Jan. 2001).
\newblock \href {https://doi.org/10.17487/RFC3031}
  {\path{doi:10.17487/RFC3031}}.
\newline\urlprefix\url{https://www.rfc-editor.org/info/rfc3031}

\bibitem{Morell2013Label}
A.~Morell, X.~Vilajosana, J.~L. Vicario, T.~Watteyne, Label switching over
  ieee802.15.4e networks, Transactions on Emerging Telecommunications
  Technologies 24 (2013).

\bibitem{rfc2205}
R.~T. Braden, L.~Zhang, S.~Berson, S.~Herzog, S.~Jamin,
  \href{https://www.rfc-editor.org/info/rfc2205}{{Resource ReSerVation Protocol
  (RSVP) -- Version 1 Functional Specification}}, RFC 2205 (Sep. 1997).
\newblock \href {https://doi.org/10.17487/RFC2205}
  {\path{doi:10.17487/RFC2205}}.
\newline\urlprefix\url{https://www.rfc-editor.org/info/rfc2205}

\bibitem{municio_simulating_2019}
E.~Municio, G.~Daneels, M.~Vučinić, S.~Latré, J.~Famaey, Y.~Tanaka, K.~Brun,
  K.~Muraoka, X.~Vilajosana, T.~Watteyne,
  \href{https://onlinelibrary.wiley.com/doi/abs/10.1002/ett.3494}{Simulating
  6tisch networks}, Transactions on Emerging Telecommunications Technologies
  30~(3) (2019) e3494.
\newblock \href {https://doi.org/10.1002/ett.3494}
  {\path{doi:10.1002/ett.3494}}.
\newline\urlprefix\url{https://onlinelibrary.wiley.com/doi/abs/10.1002/ett.3494}

\bibitem{sunori2017dead}
S.~K. Sunori, P.~K. Juneja, M.~Chaturvedi, J.~Mittal, Dead time compensation in
  sugar crystallization process, in: Proceeding of International Conference on
  Intelligent Communication, Control and Devices, Springer, 2017, pp. 375--381.

\bibitem{lim1989generalized}
K.~Lim, K.~Ling, Generalized predictive control of a heat exchanger, IEEE
  Control Systems Magazine 9~(6) (1989) 9--12.

\bibitem{leonardi2023mrt}
L.~Leonardi, L.~L. Bello, G.~Patti, Mrt-lora: A multi-hop real-time
  communication protocol for industrial iot applications over lora networks,
  Computer Communications 199 (2023) 72--86.

\bibitem{chang2016llsf}
T.~Chang, T.~Watteyne, Q.~Wang, X.~Vilajosana, Llsf: Low latency scheduling
  function for 6tisch networks (2016) 93--95.

\bibitem{kotsiou2020ldsf}
V.~Kotsiou, G.~Z. Papadopoulos, P.~Chatzimisios, F.~Theoleyre, Ldsf:
  Low-latency distributed scheduling function for industrial internet of
  things, IEEE internet of things journal 7~(9) (2020) 8688--8699.

\end{thebibliography}





\end{document}